\documentclass[a4paper,11pt]{article}
\pdfoutput=1
\usepackage[utf8x]{inputenc}
\usepackage{jheppub}

\usepackage[T1]{fontenc} 

\usepackage{amstext,amssymb}
\usepackage{amsmath}
\usepackage{graphicx}
\usepackage{xspace}
\usepackage{color}
\usepackage{units}
\usepackage{multirow}
\usepackage{slashed} 
\usepackage{comment}
\usepackage{subfigure}
\usepackage[]{hyperref}
\usepackage{appendix}
\usepackage{url}

\def \vec#1{{\boldsymbol{#1}}}

\newcommand{\bea}{\begin{eqnarray}}
\newcommand{\eea}{\end{eqnarray}}

\title{\boldmath Dark Matter in the Alternative Left Right Model}
\author[a]{\small \hspace*{-0.5cm} Mariana Frank}
\author[b,c]{\small , Chayan Majumdar}
\author[d]{\small , Poulose Poulose}
\author[b,e]{\small , Supriya Senapati}
\author[b]{\small , Urjit A. Yajnik}
\affiliation[a]{Department of Physics,  Concordia University, 7141 Sherbrooke St. West, Montreal, Quebec, Canada H4B 1R6}
\affiliation[b]{Department of Physics, Indian Institute of Technology Bombay, Powai, Mumbai, Maharashtra 400 076, India}
\affiliation[c]{Department of Physics, Middle East Technical University, TR06800, Ankara, Türkiye}
\affiliation[d]{Department of Physics, Indian Institute of Technology Guwahati, Assam 781 039, India}
\affiliation[e]{Amherst Center for Fundamental Interactions, Department of Physics,
University of Massachusetts, Amherst, MA 01003, USA}
\emailAdd{mariana.frank@concordia.ca}
\emailAdd{chayan@phy.iitb.ac.in}
\emailAdd{poulose@iitg.ac.in}
\emailAdd{ssenapati@umass.edu}
\emailAdd{yajnik@iitb.ac.in}
\abstract{
The Alternative Left-Right Model is an attractive variation of the usual Left-Right Symmetric Model because it avoids flavour-changing neutral currents, thus allowing the additional Higgs bosons in the model to be light. We show here that the model predicts several dark matter candidates naturally, through introduction of an $R$-parity similar to the one in supersymmetry, under which some of the new particles are odd, while all the SM particles are even. Dark matter candidates can be fermionic or bosonic. We present a  comprehensive investigation of all possibilities. We analyze and restrict the parameter space where relic density, direct and indirect detection bounds are satisfied,  and investigate the possibility of observing fermionic and bosonic dark matter signals at the LHC. Both the bosonic and fermionic candidates provide promising signals, the first in LHC at 300 fb$^{-1}$, the second at higher luminosity, 3000 fb$^{-1}$. Signals from bosonic candidates  are indicative of the presence of exotic $d^\prime$ quarks, while fermionic candidates imply the existence of charged Higgs bosons, all with masses in the TeV region.}

\keywords{ALRM, scotino and scalar dark matter, collider signatures.}
\begin{document} 
\maketitle
\flushbottom
\section{Introduction}
\label{sec:intro}
 
 While the discovery, 10 years ago, of the Higgs boson has provided the Standard Model (SM) with the needed missing piece, questions remain as to how complete a description of nature it offers. Thus far, its agreement with experiment is remarkable, but still outstanding questions remain, both from the theoretical foundation and the missing experimental pieces. It is hoped that looking for explanations will lead to a more complete scenario, which is the goal of Beyond The Standard Model (BSM) explorations.
 
 Although there are many phenomena unexplained in the SM, perhaps none is more mysterious than dark matter. Found to comprise about 27\% of  the universe, dark matter interacts gravitationally, and perhaps weakly, with the SM particles. The primary (particle physics) candidate is a weakly interacting massive particle (WIMP), which has not been observed, but whose indirect indications come from gravitational lensing \cite{Bartelmann:1999yn},  the cosmic microwave radiation background \cite{Lewis:2002ah,Planck:2019nip}, and other astrophysical observations. There are few models which include a natural candidate for WIMP, most notable among them is the supersymmetric extension of the SM. In supersymmetry, $R$-parity is a symmetry that distinguishes between ordinary versus supersymmetric particles. In that case, $R$-parity is a $Z_2$ symmetry defined as
 $$R=(-1)^{3B+L+2s}$$
 with $B$ the baryon number, $L$ the lepton number and $s$ the spin. Forbidding $R$-parity violation yields a stable  lightest supersymmetric particle (LSP), which is a  natural dark matter candidate.
 Other BSM scenarios deal with the existence of dark matter by introducing an {\it ad-hoc} particle, chosen as scalar, fermion, or vector. By requiring the interactions of such particle with the SM matter to obey constraints from relic abundance of dark matter \cite{Planck:2018vyg}, direct \cite{XENON:2018voc, LUX:2016ggv} and indirect  \cite{PandaX-II:2017hlx, PhysRevD.100.022001,PhysRevLett.118.251301} detection experiments, as well as collider searches, one can determine its properties (usually mass and couplings) and restrict the parameter space of the models. 
 
 In this work we shall analyze a non-supersymmetric model where  one can define a symmetry analogous to the $R$-parity in supersymmetry, yielding a natural dark matter particle. This is the Alternative Left-Right Model (ALRM). Based on the gauge group $SU(3)_C \otimes SU(2)_L \otimes SU(2)_R' \otimes U(1)_{B-L}$, the model emerges from the breaking of the exceptional group, $E_6$~\cite{Gursey:1975ki, Achiman:1978vg, Langacker:1998tc},  and it differs from the usual Left-Right Symmetric Model (LRSM) \cite{Pati:1974yy,Mohapatra:1974gc,Senjanovic:1975rk,Mohapatra:1977mj} in its assignment of right-handed fermion doublets. The ALRM avoids the unwanted tree-level flavour-changing interactions which conflict with the observed
properties of the $K$ and $B$-meson systems, which are forcing the right-handed sector of the LRSM into a high TeV range. This is avoided by the ALRM, while at the same time maintaining some of the attractive properties of the LRSM, such as providing an understanding of parity violation and a mechanism to give neutrino masses. 

The dark matter in the left-right symmetric model was analyzed most recently in \cite{Chowdhury:2021kzl}. The ALRM model has been studied before \cite{Babu:1987kp,Ma:2010us,Frank:2019nid,Frank:2020oqi}, including tests of a fermionic candidate for dark matter. However, while the notion of an extended $R$-parity was known, it was not properly explored in the context of an analysis of the dark matter sector.  We remedy this here. We show that the particles in the model can divided naturally into $R$-parity odd and $R$-parity even sectors, which do not mix with each other. The lightest $R$-parity odd particle can be a scalar, a fermion or both. We analyze here all possibilities. We impose the constraints from relic density,  direct and indirect detection experiments (to restrict the parameter space), before assessing the signatures at colliders for the different options. 

Our work is organized as follows. In Sec. \ref{sec:mod}  we present a brief discussion of the model, with emphasis on the particle structure, symmetry breaking and fermion masses. We then proceed to analyze the possibilities for dark matter candidates and their properties in Sec. \ref{sec:DM}. We look at the consequences of scalar dark matter candidates in Sec. \ref{subsec:scalarDM}, fermionic dark matter in Sec. \ref{subsec:scotinoDM}, and degenerate fermion-scalar dark matter in Sec. \ref{subsec:degenerateDM}. We investigate the possibilities of observing and of discriminating among the scenarios at the LHC in Sec. \ref{sec:collider}. Finally we summarize our findings and conclude in Sec. \ref{sec:conclusion}.

\section{The Model}
\label{sec:mod}
The particle content of the ALRM, together with their quantum number assignments under the gauge group $SU(3)_C \otimes SU(2)_L \otimes SU(2)_R' \otimes U(1)_{B-L}$ is listed in Table \ref{tab_ALRM} \cite{Frank:2019nid}.

\begin{table}[htb]
\centering
\small
\begin{tabular}{|c|c|c|c|c|c|}
\hline \hline
   Particles & $SU(3)_C$ & $SU(2)_L$ & $SU(2)_{R'}$ & $U(1)_{B-L}$ & $U(1)_{S}$          
\\[2mm]
\hline
\hline
Quarks & & & & & 
\\
\hline 
 $Q_{L} = 
\begin{pmatrix}
u_L \\
d_L
\end{pmatrix}$ & 3 & 2 & 1 & $\frac{1}{6}$ & 0   
\\[2mm]       
          $Q_{R} = 
\begin{pmatrix}
u_R \\
d'_R
\end{pmatrix}$ & 3 & 1 & $2$ & $\frac{1}{6}$ & $-\frac{1}{2}$
\\[2mm]
$d'_L$ & 3 & 1 & 1 & $-\frac{1}{3}$ & $-1$ 
\\[2mm]
$d_R$ & 3 & 1 & 1 & $-\frac{1}{3}$ & 0
\\[2mm]
\hline
Leptons & & & & & 
\\
\hline 
 $L_{L} = 
\begin{pmatrix}
\nu_L \\
e_L
\end{pmatrix}$ & 1 & 2 & 1 & $-\frac{1}{2}$ & 1
  \\[2mm]
 $L_{R} = 
\begin{pmatrix}
n_R \\
e_R
\end{pmatrix}$  & 1 & 1 & 2 & $-\frac{1}{2}$ & $+\frac{3}{2}$
          \\[2mm]
 $n_L$ & 1 & 1 & 1 & 0 & 2
          \\[2mm]
 $\nu_R$ & 1 & 1 & 1 & 0 & 1
          \\[2mm]
\hline
Scalars &&&&&    \\
\hline 
$\Phi = 
\begin{pmatrix}
\phi_1^0 & \phi_1^+\\
 \phi_2^- & \phi_2^0
\end{pmatrix}$ & 1 & 2 & $ 2^{\ast} $ & 0 & $-\frac{1}{2}$ 
          \\[2mm]
         $\chi_{L} = 
\begin{pmatrix}
\chi_L^+ \\
\chi_L^0
\end{pmatrix}$ & 1 & 2 & 1 & $\frac{1}{2}$ & 0 
        \\[2mm]
         $\chi_{R} = 
\begin{pmatrix}
\chi_R^+ \\
\chi_R^0
\end{pmatrix}$ & 1 & 1 & 2 & $\frac{1}{2}$ & $\frac{1}{2}$ 
        \\[2mm]          
  \hline
  Gauge Bosons &&&&& \\
  \hline
   $B^\mu$ & 1 & 1 & 1 & 0 & 0 
   \\[2mm]
    $W^\mu_L$ & 1 &3 & 1 &0 & 0
    \\[2mm]
    $W^\mu_R$ & 1 & 1 & 3 & 0 & 0
    \\[2mm]
    \hline
    \hline
\end{tabular}
\caption{\small Particle content of ALRM together with their quantum numbers under $SU(3)_C \otimes SU(2)_L \otimes SU(2)_{R'} \otimes U(1)_{B-L} \otimes U(1)_{S}$ considered in this study. While several $U(1)_S$ assignments exist in the literature, we follow \cite{Frank:2019nid}, and define the generalized lepton number  as $L=S+T_{3R'}$. }
\label{tab_ALRM}
\end{table}

As they emerge from different choices of breaking $E_6$, the quark and lepton assignments in multiplets differ from those in the LRSM. Here the left-handed (LH) fermions form $SU(2)_L$ doublets, while the right-handed (RH) up-type quarks form doublets with exotic down-type quarks $d_R^\prime$. Similarly, while the LH leptonic doublets are the same as in the SM, the RH charged leptons partner with exotic RH neutral fermions ($n_R$) to form $SU(2)_{R^\prime}$ doublets. In addition, there are the left-handed partners of exotic quarks ($d_L'$), right-handed partners of the usual quarks  ($d_R$), and two neutral fermions, $n_L$ and $\nu_R$, which are singlets under both $SU(2)_L$ and $SU(2)_{R'}$ \cite{Frank:2019nid, Ashry:2013loa,Ashrythesis}.  
The electroweak symmetry breaking occurs as follows. First, the gauge and global
symmetry $SU(2)_{R'} \times U(1)_{B-L} \times U(1)_S$ is broken  to
the hypercharge $U(1)_Y$ while preserving the generalized lepton number defined as $L=S+T_{3R'}$.
This is achieved through the vacuum expectation value (VEV) of the $SU(2)_{R'}$ doublet  scalar $\chi_R$, which is 
charged under $U(1)_S$. The electroweak symmetry is then broken down to
electromagnetism by means of the VEV of the bidoublet  Higgs field $\Phi $, charged under both
$SU(2)_L$ and $SU(2)_{R'}$, but with no $U(1)_{B-L}$ quantum numbers, as well as by the VEV of $\chi_L$ \cite{Khalil:2009nb,Khalil:2010yt,Ma:2012gb,Frank:2004vg, Frank:2019nid}.

The model Lagrangian includes  standard gauge-invariant kinetic terms
for all fields, a Yukawa interaction Lagrangian ${\cal L} _{\rm Y}$,  and the scalar
potential $V_{\Phi \chi}$. The most general Yukawa Lagrangian respecting  gauge
and  global $U(1)_S$ symmetries is 
 \begin{equation}
\label{eq:yuk}
 { \cal L} _{\rm Y} = \bar Q_L {\bf \hat Y}^u \hat\Phi Q_R
    - \bar Q_L {\bf \hat Y}^d_L \chi_L d_R
    - \bar Q_R {\bf \hat Y}^{d'}_R \chi_R d'_L
    - \bar L_L {\bf \hat Y}^e \Phi L_R
    + \bar L_L {\bf \hat Y}^\nu_L \hat \chi_L \nu_R
    + \bar L_R {\bf \hat Y}^n_R \hat \chi_R n_L + {\rm h.c.} \ ,
\end{equation}
 with flavour indices omitted for clarity, where the Yukawa
couplings ${\bf \hat Y}$ are $3 \times 3$ matrices in the flavour space.
The hatted quantities refer to the duals of the scalar fields
$\hat\Phi=\sigma_2\Phi\sigma_2$ and $\hat\chi_{L,R}= i \sigma_2 \chi_{L,R}$.
Unlike conventional LRSM, in this framework right-handed neutrinos are gauge singlets which  allows us to introduce a bare Majorana mass term,
\begin{equation}
\mathcal{L}_M = m_M \bar{\nu_R}^c \nu_R.
\label{eq:LM}
\end{equation}
The most general Higgs potential $V_{\Phi \chi}$ preserving  left-right symmetry is ~\cite{Borah:2010zq}
\begin{eqnarray}
  V_{\Phi \chi} &= &
   -\mu_1^2 {\rm Tr} \big[ \Phi^\dag \Phi\big]
   -\mu_2^2 \big[\chi_L^\dag \chi_L + \chi_R^\dag \chi_R\big]
   + \lambda_1 \big({\rm Tr}\big[\Phi^\dag \Phi\big]\big)^2
   + \lambda_2\ \rm{Tr} (\hat\Phi^\dag \Phi)\rm{Tr}(\Phi^\dag \hat\Phi)  \nonumber\\
   &+&\lambda_3 \Big[\big(\chi_L^\dag \chi_L\big)^2 +
         \big(\chi_R^\dag\chi_R\big)^2\Big] 
+ 2 \lambda_4\ \big(\chi_L^\dag \chi_L\big)\ \big(\chi_R^\dag\chi_R\big)
   + 2 \alpha_1 {\rm Tr} \big[\Phi^\dag \Phi\big]
          \big[\chi_L^\dag \chi_L + \chi_R^\dag \chi_R\big] \nonumber \\
          &+& 2 \alpha_2 \big[ \big(\chi_L^\dag \Phi\big) \big(\Phi^\dag \chi_L\big) +\big(\chi_R^\dag \Phi^\dag \big)\ \big(\Phi\chi_R\big)\big] 
+ 2 \alpha_3 \big[ \big(\chi_L^\dag \hat\Phi\big)\
          \big(\hat\Phi^\dag \chi_L \big) + \big(\chi_R^\dag \hat\Phi^\dag \big)\
          \big(\hat\Phi \chi_R\big)\big] \nonumber \\
&+& \mu_3 \big[\chi_L^\dag \Phi \chi_R + \chi_R^\dag\Phi^\dag\chi_L\big] \ ,
\label{eq:Hpot}
\end{eqnarray}
The properties of this potential and implications for the vacuum stability of the model have been analyzed in \cite{Frank:2020oqi}.
After the breaking of the left-right symmetry down to $U(1)_{\rm em}$, the
neutral components of the scalar fields acquire non-vanishing VEVs,
\begin{equation} 
  \langle \Phi  \rangle = \frac{1}{\sqrt{2}}\begin{pmatrix} 0&0\\0 & v_2 \end{pmatrix}\ , \qquad
  \langle \chi_L\rangle = \frac{1}{\sqrt{2}}\begin{pmatrix} 0\\ v_L \end{pmatrix}\ , \qquad
  \langle \chi_R\rangle = \frac{1}{\sqrt{2}}\begin{pmatrix}0\\ v_R \end{pmatrix} ,
\label{eq:symbreak}
\end{equation}
with the exception of $\phi^0_1$, which is protected by the conservation of the
generalized lepton number, which also forbids mixing between the SM $d$ and
exotic $d^\prime$ quarks.

Left-right symmetry breaking generates masses for the
model gauge bosons,  and  Higgs-boson kinetic
terms are responsible for their mixing. Because  $\langle\phi_1^0
\rangle = 0$, the charged $W=W_L$ and $W'=W_R$ bosons do not mix, and their masses are given by
\begin{equation}
  M_{W_L}    = \frac12 g_L \sqrt{v_2^2+v_L^2} \equiv \frac12 g_L v
 \qquad\text{and}\qquad
  M_{W_R} = \frac12 g_R \sqrt{v_2^2+v_R^2} \equiv \frac12 g_R v' \ ,
\label{eq:mw_mwp}
\end{equation}
with $g_L$ and $g_R$ the coupling constants for $SU(2)_L$ and $SU(2)_{R'}$. This releases the $W_R$ boson from contributing to low energy physics, in particular to flavor violation in $K^0-{\bar K}^0$ and $B^0-{\bar B}^0$ (as $W_R$ always couples to $ud^\prime$) and lifts all constrains on its mass. However, the $W_R$ is related to $Z_R$ mass. In the neutral sector, the gauge boson squared mass matrix is written, in the
$(B_\mu, W_{L\mu}^3, W_{R\mu}^3)$ basis, as
\begin{equation}
  ({\cal M}^0_V)^2 = \frac14 \begin{pmatrix}
    g_{B-L}^2\ (v_L^2+v_R^2)  & -g_{B-L} g_L v_L^2    & -g_{B-L}\ g_R\ v_R^2\\
   -g_{B-L}\ g_L\ v_L^2       &  g_L^2\ v^2          & -g_L\ g_R\ v_2^2\\
   -g_{B-L} \ g_R \ v_R^2       & -g_L\ g_R\ v_2^2       & g_R^2\ v^{\prime 2}
  \end{pmatrix}\, ,
\end{equation}
with $g_{B-L}$ the coupling constant for $U(1)_{B-L}$. The matrix can be diagonalised through three rotations that mix the $B$, $W_L^3$ and
$W_R^3$ bosons into the massless photon $A$ and massive $Z$ and $Z'$ states.
Neglecting the $Z$/$Z'$ mixing, the $Z$ and $Z'$ boson
masses are given by
\begin{eqnarray}
  M_{Z} =  \frac{g_L}{2 \cos \theta_W} \ v
  \quad\text{and}\quad
  M_{Z^\prime} = \frac12 \sqrt{\frac {g_{B-L}^4( v_L^2 + v_R^2)+g_R^4( v_2^2 + v_R^2)+2g_{B-L}^2g_R^2 v_R^2}{g_{B-L}^2+g_R^2}}\, ,
\label{eq:mz_mzp}
\end{eqnarray}
where $ \theta_W$-rotation denotes the usual electroweak mixing with the definition
$\text{sin}\theta_W = \frac{g_Y}{\sqrt{g_L^2 + g_Y^2}} = \frac{e}{g_L}$
where $e$ and $g_Y$ denote electromagnetic coupling constant and hypercharge respectively \cite{Frank:2019nid}. The relationship between $M_{W_R}$ and $M_{Z^\prime}$ is not as transparent as in LRSM. However, unlike $W_R$, $Z^\prime$ couples to ordinary quarks and leptons, and its mass is restricted by measurements at ATLAS \cite{ATLAS:2017fih}, requiring $v_R$ to be large. In the limits $v_R \gg v_2, v_L$
\begin{equation}
\frac {M_{Z^\prime}}{M_{W_R}} \sim \frac{\sqrt{g_{B-L}^2+g_R^2}}{g_R} \, ,
\end{equation}
forcing $W_R$ to be heavy\footnote{In fact the ratio $\frac {M_{Z^\prime}}{M_{W_R}} $ is lower in ALRM than in the LRSM, meaning that the spectrum is more compressed in the latter \cite{Frank:2019nid}.}.

Fermion masses  are generated from the Yukawa Lagrangian (\ref{eq:yuk}),
 after the breaking of the $SU(2)_L\times SU(2)_{R'}\times
U(1)_{B-L}$ symmetry down to $U(1)_{\rm em}$.
The resulting fermion masses are
\begin{eqnarray}
m_u=\frac{1}{\sqrt{2}}Y^u v_2 \sin \beta \,, \quad m_d=\frac{1}{\sqrt{2}}Y^d_L v_2 \cos \beta \,, \quad m_{d'}=\frac{1}{\sqrt{2}}Y^{d^\prime}_R v_R \,, \quad
m_e=\frac{1}{\sqrt{2}}Y^e v_2 \sin \beta  \,, \nonumber\\
\end{eqnarray} 
with $\tan\beta=v_2/v_L$. Allowing for soft-breaking of the lepton number, $L$,  $\nu_R$ can acquire a Majorana mass as given in Eq. \ref{eq:LM}, and the consequent lepton asymmetry could lead to leptogenesis \cite{Frank:2020odd}.
 The Dirac mass for the left-handed, Majorana mass for the right-handed neutrinos and Dirac mass for scotinos are, respectively
\begin{equation}
M_\nu=\frac{1}{\sqrt{2}}Y^\nu_L v_L
\,, \quad M_N = m_M \,, \quad M_n=\frac{1}{\sqrt{2}}Y^n_R v_R\, ,
\end{equation}
We assume neutrinos to be Majorana particles. Defining 
\begin{equation}
\nu=\frac{\nu_L+\nu_L^c}{\sqrt{2}}\, , \qquad N=\frac{\nu_R+\nu_R^c}{\sqrt{2}}\, ,
\end{equation}
light left-handed neutrino masses are generated through the see-saw mechanism
\begin{equation}
\begin{pmatrix} {\bar \nu_L} & {\bar \nu_R}^c \end{pmatrix}
\begin{pmatrix} 0 & M_\nu \\M_\nu^T & M_N \end{pmatrix}
\begin{pmatrix} \nu_L^c \\ \nu_R \end{pmatrix} \,.
\end{equation}
Here $\nu$ and $N$ are approximate eigenstates with masses
\begin{equation}
m_N \simeq M_N \qquad {\rm and} \qquad m_\nu= M_\nu M_N^{-1} M_\nu^T\, ,
\end{equation}
with $M_N$ assumed to be diagonal, and $m_\nu$ is diagonalized in flavor eigenstates as 
\begin{equation}
m_\nu^{\rm diag}= V_L^{\nu \ \dagger} m_\nu V_L^\nu \, .
\end{equation}

  Other details of the model, including a complete description of the gauge sector, can be found in Refs. \cite{Ashry:2013loa,Ashrythesis,Frank:2004vg, Borah:2010zq, Frank:2020odd}.  


 \section{Dark Matter}
 \label{sec:DM}
 
The ALRM augmented by the extra $U(1)_S$ symmetry allows  the introduction of the generalized lepton number $L=S+T_{3R}$. Similarly one can introduce a generalized $R$-parity, similar to the one existing in supersymmetry,  defined here in a similar way as $(-1)^{3B+L+2s}$  \cite{Khalil:2009nb}.  Under this $R$-parity,  all SM quarks, leptons and SM gauge bosons are even. The odd $R$-parity particles are as follows: in the scalar sector, $\chi_R^\pm, \, \phi_1^\pm, \, \Re(\phi_1^0)$ and $\Im(\phi_1^0)$, in the fermion sector, the scotinos $n_L, n_R$, and the exotic quarks $d^\prime_L, d^\prime_R$, and in the gauge sector,  $W_R$.  All the rest of the particles in the spectrum are $R$-parity even. The existence of a dark matter sector arising from $R$-parity odd particles is another attractive feature of this model, and an advantage over the more common LRSM.  
 Of the $R$-parity odd particles, only the neutral Higgs, the pseudoscalar Higgs or the scotino can be DM candidates, as the rest of the particles are electromagnetically charged, and in addition, $d^\prime$ quarks have strong interactions.   Thus, in what follows we shall investigate the possibility that the DM is either the $R$-parity odd Higgs boson (scalar or pseudoscalar), or  the scotino(s), or both.\footnote{ The possibility that the scotino is a dark matter candidate, without the association to $R$-parity,  was examined in \cite{Frank:2004vg, Frank:2019nid, Ma:2010us, Khalil:2009nb, Khalil:2010yt, Ma:2012gb}.} We proceed to analyze these in turn, looking for ways to observe and distinguish them.

 \subsection{Scalar Dark Matter}
 \label{subsec:scalarDM}
 
The scalar sector of the model consists of one bidoublet $\Phi$ Higgs field,  with the $SU(2)_L$ symmetry acting along the columns and the $SU(2)_{R^\prime}$ along the rows, as  in Table~\ref{tab_ALRM}.  In addition to the bidoublet Higgs boson $\Phi$ there are two doublet Higgs fields transforming  under $SU(2)_L$ and $SU(2)_{R^\prime}$, denoted by $\chi_{L}$ and $\chi_R$, respectively.  $SU(2)_{R^\prime}\otimes U(1)_{B-L}$ to $U(1)_Y$ breaking is induced by the VEV of $\chi_R$, while  the electroweak symmetry breaking is driven  by the VEVs of $\Phi$ and $\chi_L$. 
The global $U(1)_S$ symmetry insures that the quark doublets can interact with $\hat{\Phi}$ and lepton doublets with $\Phi$ only. This symmetry also restricts $\phi_1^0$ from acquiring a VEV, forbidding  the $W_L-W_R$ gauge boson mixing, as well as the $d - d^\prime$ and $n- \nu$ mixing in the model. 
The conservation of $R$-parity as defined above is reflected in the Higgs boson mixing matrices, which are consistent with the absence of the mixing between the $R$-parity even and $R$-parity odd scalars. In the charged scalar sector, the squared mass matrix is block diagonal. The $\phi^\pm_2$ and $\chi_L^\pm$ ($R$-parity even) fields  mix  independently of the $\phi_1^\pm$ and $\chi_R^\pm$ ($R$-parity odd) fields. The  $2\times 2$ block  mass matrices
$({\cal M}^\pm_L)^2$ and $({\cal M}^\pm_R)^2$ are, in
the $(\phi_2^\pm, \chi_L^\pm)$ and $(\phi_1^\pm, \chi_R^\pm)$ bases, respectively, as
\renewcommand{\arraystretch}{1.4}
\begin{eqnarray}
  ({\cal M}^\pm_{L,R})^2 = \begin{pmatrix}
     -(\alpha_2-\alpha_3)v_{\scriptscriptstyle L,R}^2 -
         \frac{\mu_3 v_L v_R}{\sqrt{2}v_2}~~~~~ & 
     (\alpha_2-\alpha_3) v_2 v_{\scriptscriptstyle L,R} +
         \frac{\mu_3 v_{\scriptscriptstyle R,L}}{\sqrt{2}}\\
     (\alpha_2-\alpha_3) v_2 + \frac{\mu_3 v_{\scriptscriptstyle R,L}}{\sqrt{2}}&
     -(\alpha_2-\alpha_3)v_2^2 - \frac{\mu_3 v_2 v_{\scriptscriptstyle R,L}}
       {\sqrt{2}v_{\scriptscriptstyle L,R}}\end{pmatrix} \ ,
\end{eqnarray}

The masses of the charged Higgs bosons  are obtained  by diagonalizing the $2 \times 2$ matrices: 
\begin{eqnarray}
    m^2_{H_1^{\pm}} &=& - \left[ v_2 v_L \left( \alpha_2 - \alpha_3 \right) + \frac{\mu_3 v_R}{\sqrt{2}}\right]\frac{v^2}{v_2 v_L}\\
    \label{charged1}
    m^2_{H_2^{\pm}} &= &- \left[ v_2 v_R \left( \alpha_2 - \alpha_3 \right) + \frac{\mu_3 v_L}{\sqrt{2}}\right]\frac{v^{\prime 2}}{v_2 v_R}
    \label{charged2}
\end{eqnarray}
with $v^2 = v_2^2 + v_L^2$ and $v^{\prime 2} = v_2^2 + v_R^2$. Here $H_2^\pm$ is $R$-parity odd and $H_1^\pm$ is $R$-parity even.
The other two eigenstates of these matrices correspond to the Goldstone bosons $G_2^\pm$ ($R$-parity even) responsible for giving mass to the $W_L^\pm$ boson, and the Goldstone $G_1^\pm$ ($R$-parity odd) giving mass to $W_R^\pm$,  also odd under $R$-parity.

In the neutral scalar sector, components of the $\phi_1^0$ field ($\Re(\phi_1^0)$ and $\Im(\phi_1^0)$)  do not mix with other states, as they are both $R$-parity odd. They yield the  physical $H_1^0$ and $A_1$ eigenstates, which are degenerate in mass. Calling these masses   $M_{H_1^0}$ and $M_{A_1}$, we obtain
\begin{eqnarray}
    M_{H_1^0}^2=M_{A_1}^2 &=& 2 v_2^2 \lambda_2 - \left( \alpha_2 - \alpha_3 \right) \left(v_L^2 + v_R^2 \right) - \frac{\mu_3 v_L v_R}{\sqrt{2}v_2}.
    \label{odd1}
    \end{eqnarray}
   Thus, if these are the lightest $R$-parity odd particles, $H_1^0$ and $A_1$ would form the two components  of  the dark matter.  
  
For the
three remaining scalar and pseudoscalar fields (all of which are $R$-parity even), the mass-squared matrices  $({\cal M}^0_\Re)^2$ and $({\cal M}^0_\Im)^2$ of  in the
$(\Re\{\phi^0_2\}$, $\Re\{\chi_L^0\}$, $\Re\{\chi_R^0\})$ and
$(\Im\{\phi^0_2\}$, $\Im\{\chi_L^0\}$, $\Im\{\chi_R^0\})$ bases, are given, respectively, by 
\renewcommand{\arraystretch}{1.3}
\setlength\arraycolsep{5pt}
\begin{eqnarray}
  ({\cal M}^0_\Re)^2 &=&\ \begin{pmatrix}
    2v_2^2 \lambda_1 \!-\! \frac{\mu_3 v_L v_R}{\sqrt{2} v_2} & 2 \alpha_{12} v_2 v_L
      \!+\! \frac{\mu_3 v_R}{\sqrt{2}} & 2 \alpha_{12} v v_R \!+\!
      \frac{\mu_3 v_L}{\sqrt{2}}\\
    2\alpha_{12} v_2 v_L \!+\! \frac{\mu_3 v_R}{\sqrt{2}} & 2 \lambda_3 v_L^2
      \!-\!\frac{\mu_3 v_2 v_R}{\sqrt{2} v_L} & 2 \lambda_3 v_L v_R\!-\!
      \frac{\mu_3 v_2}{\sqrt{2}}\\
    2\alpha_{12} v_2 v_R \!+\! \frac{\mu_3 v_L}{\sqrt{2}} & 2\lambda_3 v_L v_R\!-\!
       \frac{\mu_3 v_2}{\sqrt{2}}  & 2 \lambda_3 v_R^2 \!-\!
       \frac{\mu_3 v_2 v_L}{\sqrt{2} v_R}
  \end{pmatrix} 
  \label{eq:cpevenH0}\ , \\
  ({\cal M}^0_\Im)^2 &=&\ \frac{\mu_3}{\sqrt{2}} \begin{pmatrix}
    -\frac{v_L v_R}{v_2} & v_R & - v_L \\
    v_R & -\frac{v_2 v_R}{v_L} & v_2\\
    -v_L & v_2 & -\frac{v_2 v_L}{v_R} \end{pmatrix} \ ,
\end{eqnarray}
where $\alpha_{12}=\alpha_1+\alpha_2$. 
The mass of the pseudoscalar boson $A_2$ is
  \begin{eqnarray}
    M_{A_2}^2 &=&  - \frac{\mu_3 v_L v_R}{\sqrt{2}v_2} \left[ 1+ v_2^2 \left(\frac{1}{v_L^2} + \frac{1}{v_R^2} \right)\right] \, .
    \label{odd2}
\end{eqnarray}
 The other two CP-odd states  in  
$({\cal M}^0_\Im)^2$ are Goldstone bosons $G_1^0$ and $G_2^0$  corresponding to the gauge bosons $Z_L$ and $ Z_R$, respectively.  Both the Goldstone bosons and neutral gauge bosons are $R$-parity even, thus conserving $R$-parity.  Furthermore, unlike in the case of the charged gauge bosons, $Z_L$ and $Z_R$ can mix, to give the mass eigenstates $Z$ and $Z'$ \cite{Ashry:2013loa,Ashrythesis}. Thus, in fact, while LHC constraints  $Z'$ mass, the $W_R$ mass cannot be constrained directly as it decays into exotic quarks. However, the mass relations between $Z'$ and $W_R$ indirectly restricts the latter through the direct constraints on the former.

In the CP-even, $R$-parity even scalar sector, we denote the mass eigenstates (the eigenstates of ${\cal M}^0_\Re$ in Eq.~\ref{eq:cpevenH0}) as 
$H^0_{0,2,3}$, and further identify the lightest state $H^0_0$ as the  SM-like Higgs boson, $h$.
It is then conventional to fix the quartic coupling, $\lambda_1$  in terms the mass of the lightest Higgs state, $M_h=125$ GeV, giving 
\begin{eqnarray}
\label{eq:lam1}
  \lambda_1 =  \frac{1}{2 v^3}\ \frac{
    \sqrt{2} v_2 v_L v_R M_{h}^6 + \mathfrak{a}^{(4)} M_{h}^4
      - 2\mathfrak{a}^{(2)}M_{h}^2 - 4\alpha_{12}^2\mu_3 v_2^4(v_L^2-v_R^2)^2
  }{
    \sqrt{2} v_L v_R M_{h}^4
      + (\mu_3 v_2 - 2\sqrt{2}\lambda_3 v_L v_R) (v_L^2+v_R^2)M_{h}^2 
      - 2\mu_3 v_2 \lambda_3 (v_L^2-v_R^2)^2
  }\ .
\end{eqnarray}
With this  masses of the other  two CP-even ($R$-parity-even) neutral Higgs boson states are given by
    \begin{eqnarray}
    M_{H^0_{2,3}}^2 &=& \frac{1}{2} \left [\mathfrak{a} - M_{h}^2 \mp \sqrt{\left(\mathfrak{a} - M_{h}^2 \right)^2 + 4 \left(\mathfrak{b} + M_{h}^2 \left(\mathfrak{a} - M_{h}^2 \right) \right) }\right]\,,
    \label{even2}
\end{eqnarray}
where 
\begin{eqnarray}
    \mathfrak{a} &= &2 v_2^2 \lambda_1 + 2 \left(v_L^2 + v_R^2 \right)\lambda_3 - \frac{\left( v_L^2 v_R^2 + v_2^2 \left[v_L^2 + v_R^2 \right)\right] \mu_3}{\sqrt{2}v_2 v_L v_R}, \nonumber \\
    \mathfrak{b} &=& \frac{\left( v_L^2 + v_R^2 \right)}{v_2 v_L v_R} \left\{ 4 v_2^3 v_L v_R \left[ \left( \alpha_1 + \alpha_2 \right)^2 -  \lambda_1 \lambda_3  \right] + \sqrt{2}v_2^4 \lambda_1 \mu_3 + \sqrt{2}v_L^2 v_R^2 \lambda_3 \mu_3 \right\} \nonumber \\
    &+& \frac{\sqrt{2}v_2 \mu_3}{v_L v_R} \left[ 4 v_L^2 v_R^2 \left( \alpha_1 + \alpha_2 \right) + \left(v_L^2 - v_R^2 \right)^2 \lambda_3 \right]. \nonumber 
\end{eqnarray}


We now proceed to investigate the consequences of having $H_1^0$ and  $A_1$  as degenerate scalar dark matter candidates. 
First, we must ensure that the chosen particles satisfy experiments searching for dark matter, and their interactions with ordinary matter. Searches for dark matter fall into three categories: (ii) direct detection, where dark matter is observed through possible scattering events with baryonic matter such as protons and neutrons making up atomic nuclei in the detectors; (ii) indirect detection, where dark matter particles annihilate into SM particles such as photons, positrons, antiprotons etc; and (iii) production at colliders, where dark matter particles are created from colliding of SM particles. (See the review \cite{Feng:2010gw}).

In direct detection, experiments aim to detect nuclear recoils emerging from collisions between dark matter particles and a detector target, such as Xenon. Discriminating the signal emerging from dark matter particles from the background requires concentrating on characteristic signatures of dark matter, such as the energy dependence or the annual modulation of the signal \cite{Drukier:1986tm}. From the non-observation of the signal, stringent bounds were imposed on the nucleon scattering cross section of the DM particle in both spin-independent (SI) and spin-dependent (SD) cross sections. Only the former is applicable here, as scalars are spinless.   Several experiments such as XENON \cite{XENON:2018voc},   LUX \cite{LUX:2016ggv}, PANDA \cite{PandaX-II:2017hlx}, and PICO \cite{PhysRevD.100.022001,PhysRevLett.118.251301} have set stringent limits on dark matter particle masses based on non-observation of signals. 

Complementary to direct detection, indirect detection of dark matter relies on detection of photons  and/or electrons or positrons as products of annihilating dark matter particles. Ideal targets are those where the dark matter density is high and the background from astrophysical processes is small or well known, such as galaxy clusters, making the Galactic Centre of the Milky Way a fertile ground for H.E.S.S \cite{Hess:2021cdp} and Fermi-LAT \cite{Karwin:2016tsw} experiments.  

Collider experiments focus on production of dark matter particles in colliders like the LHC in association with other SM particles. In such case, the DM would not be detected and become part of the missing information. This is much like the information that would lost in events with neutrinos in the final state. From the measurement of missing transverse momentum and energy, one can infer the presence of these missing particles. We shall discuss possible LHC signatures in the scenario discussed here in Sec.~\ref{sec:collider}.

To explore the viability of $H_1^0$ and  $A_1$ as dark matter candidates, we first investigate the amount of thermal relic abundance of dark matter,  aiming to restrict the parameter space based on agreement with the measured cosmological dark matter density. For this, we use the implementation of the ALRM into {\tt FeynRules}  \cite{Alloul:2013bka,Frank:2019nid}, which encodes the information on the particle content of the model and then link this to {\tt MicrOmegas} package \cite{Belanger:2018ccd}, which calculates relic abundance, and direct and indirect cross sections relevant to dark matter. 
In the early universe  $H_1^0$ and  $A_1$ would have been produced in abundance, and they were  in thermal equilibrium. However, as the Universe cooled down and the temperature came down to near the mass of the dark matter, the annihilation process became more relevant and the reactions fell out of equilibrium, resulting in a fall of the number density of the dark matter. This continued until the reaction rate is larger than the Hubble expansion parameter. Beyond this temperature, the number density of the dark matter remains the same. This freeze-out and the remaining relic abundance of the dark matter thus depends crucially on the annihilation reactions and their cross sections.  The total dark matter relic density obtained from the anisotropy in the cosmic microwave background radiation (CMBR) measured by the Planck experiment is, $\Omega^{obs}_{DM}h^2=0.120 \pm 0.001$ \cite{Planck:2018vyg}, where $h$ is the dimensionless Hubble constant.  An immediate goal of the present study is to find the parameters of the model that give the right annihilation cross section so as to obtain the required relic density.  The annihilation channels available depend on the mass of the dark matter candidates  ($H_1^0$ and $A_1$). Only channels where sum of the masses of the final states is smaller than $2M_{H_1^0}$ will contribute. 
Table \ref{tab:Hannihilation} lists all the  annihilation channels  for different  ranges of  $M_{H_1^0}=M_{A_1}$.  
We restrict our study to $M_{H_1^0}$ up to 2 TeV, which is much lighter than the $Z'$, and therefore channels with $Z'$ in the final state are irrelevant here.

\begin{table}[htb]
\begin{center}
\begin{tabular}{c|c}
 \hline\hline
        $M_{H_1^0}$ (GeV)  & Annihilation Channels \\
\hline\hline
 $M_{H_1^0}<M_h$ & $H^0_1 H^0_1 \rightarrow WW, ~ZZ, ~GG, ~\gamma \gamma, ~b\bar{b}$ \\ 
 \hline
  $M_h<M_{H_1^0}<m_t$ & $H^0_1 H^0_1 \rightarrow WW, ~ZZ,~ GG,~ \gamma \gamma, b\bar{b},~hh$ \\ 
 \hline
 $m_t<M_{H_1^0}<M_{H_1^\pm}$& $H^0_1 H^0_1 \rightarrow WW, ZZ, GG, \gamma \gamma,~b\bar b,~hh,~{t\bar t}$ \\
 \hline
$(M_{H_1^\pm}+M_W) <2 M_{H_1^0}$ & $~~~H^0_1 H^0_1 \rightarrow WW, ZZ, GG, \gamma \gamma, ~b\bar b,~hh,~{t\bar t},~W^\pm H_1^\mp$ \\
 \hline 
 \hline 
\end{tabular}
\end{center}
\caption{Annihilation channels for $H^0_1$ for different scalar dark matter masses. }
\label{tab:Hannihilation}
\end{table}

\begin{figure}[htb!]
    \centering
   \hspace{-0.5cm} \includegraphics[scale=0.36]{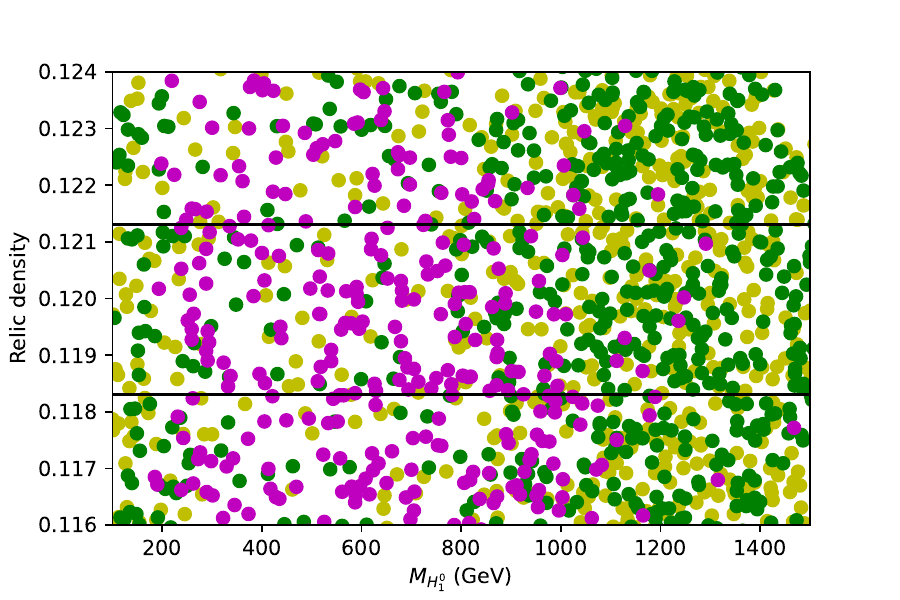}\hspace{-0.5cm}
    \includegraphics[scale=0.35]{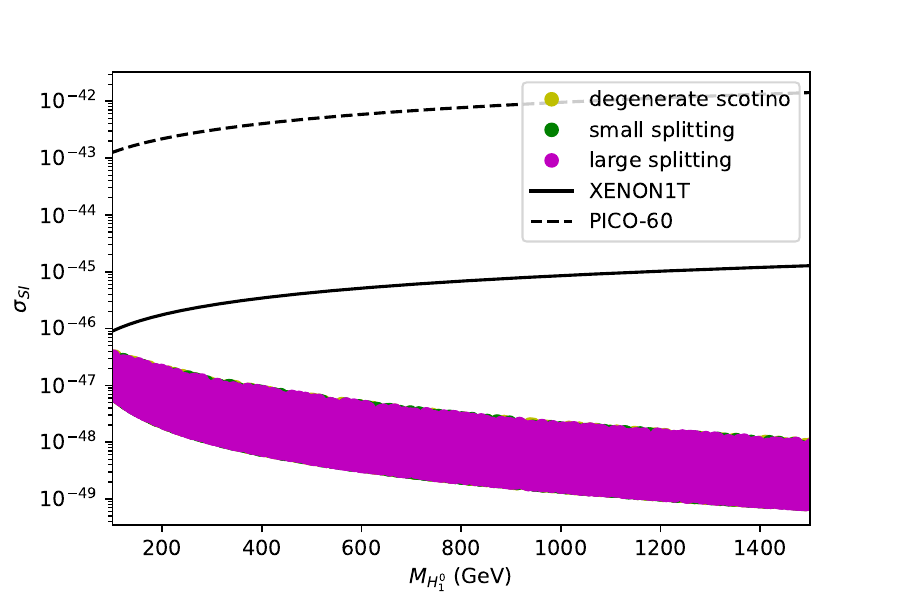}\hspace{-0.5cm}
    \includegraphics[scale=0.35]{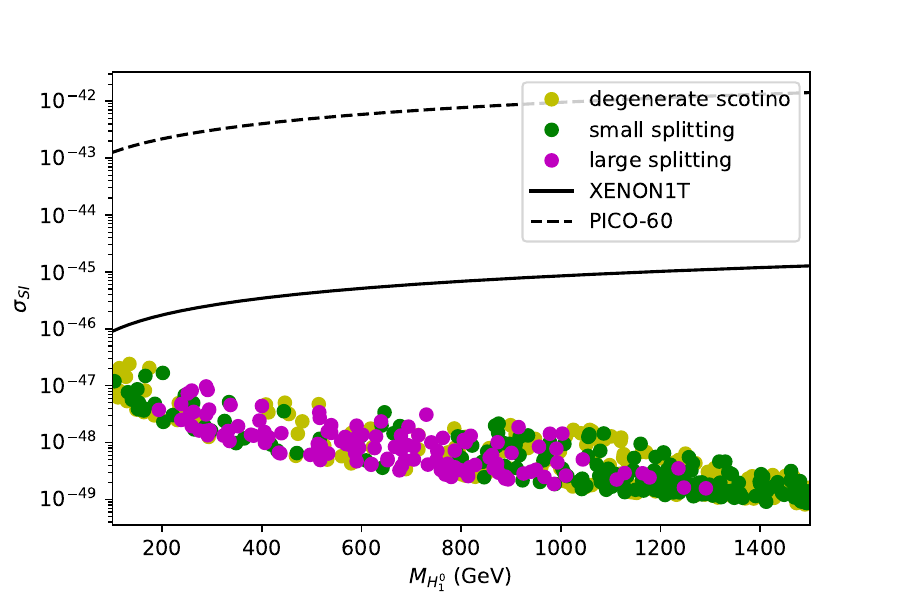}
    \caption{ (Left): Relic density vs. $M_{H_1^0}$ with horizontal black lines showing $1\sigma$ range $\Omega^{obs}_{DM}h^2$.
     (Middle): Direct detection cross section plotted against $M_{H_1^0}$, along with bounds from XENON1T (in black solid) and PICO experiments (in black dashed). As the direct detection cross sections are independent of scotino masses, all the three cases overlap.
   (Right): Direct detection cross section plotted against $M_{H_1^0}$, with points satisfied by $\Omega_{obs}$ within $1\sigma$. 
Different colours correspond to different scotino mass hierarchy with Case (i) (greenish yellow), Case (ii) (green) and Case (iii) (magenta) given in Table~\ref{tab:DMmassrange}.  }
    \label{fig:my_label}
\end{figure}

\begin{figure}[htb!]
    \centering
   \hspace{-0.5cm} \includegraphics[scale=0.36]{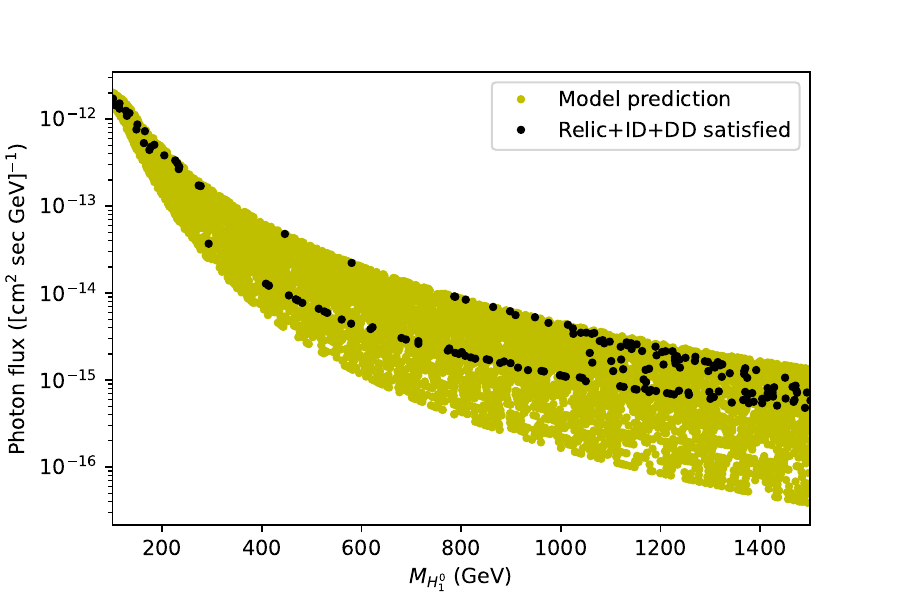}\hspace{-0.5cm}
    \includegraphics[scale=0.35]{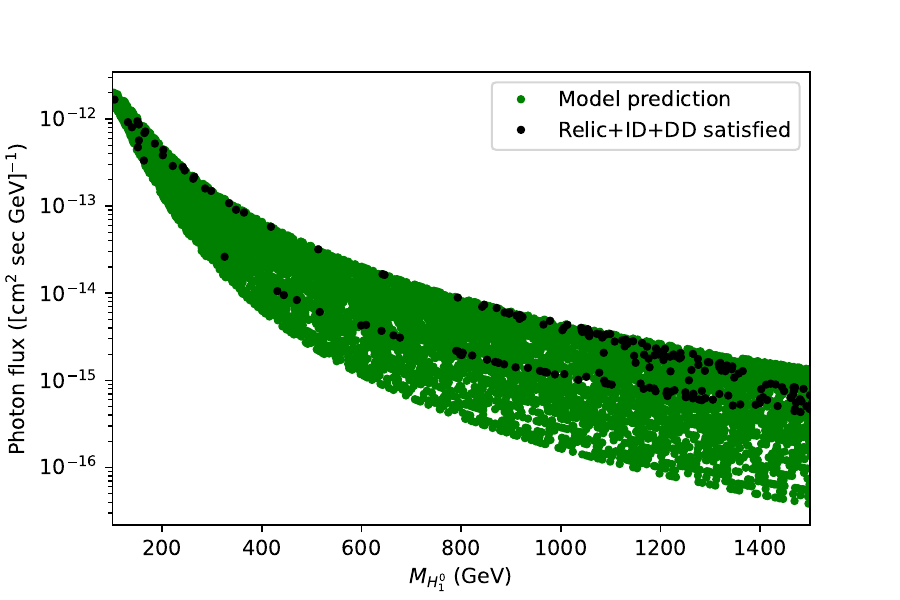}\hspace{-0.5cm}
    \includegraphics[scale=0.35]{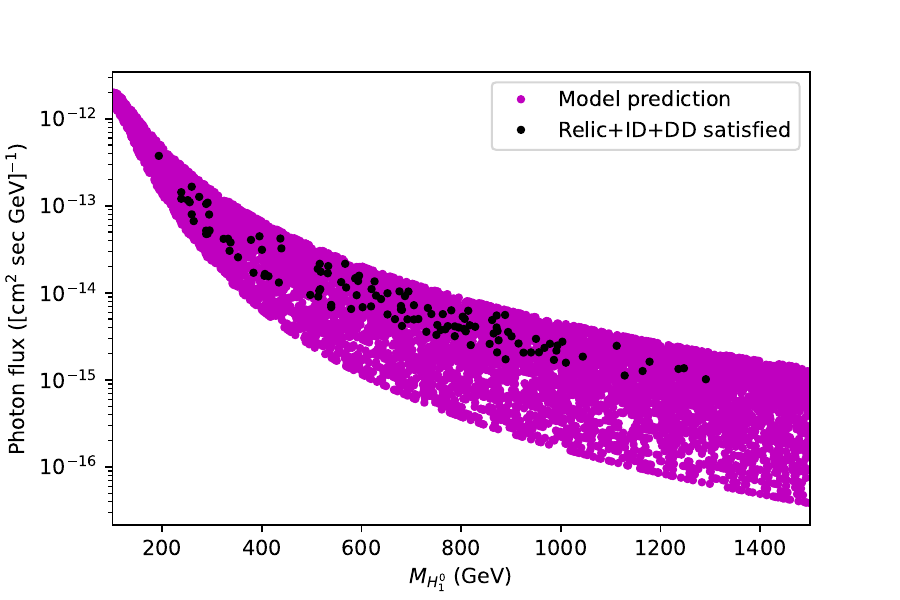}
    \caption{Indirect detection restrictions case with the photon flux versus dark matter mass for Case (i) - degenerate scotino mass scenario (Left), Case (ii) - small mass splitting (Middle), and Case (iii) - large mass splitting (Right). All points satisfy constraints from Fermi-LAT \cite{Karwin:2016tsw}, while the black dots are also consistent with relic density + DD restrictions imposed by PICO-60 as well as XENON1T experiments (Fig.~\ref{fig:my_label}).}
    \label{fig:scalarID}
\end{figure}

We consider the following independent parameters: $\alpha_1, ~\alpha_2 = \alpha_3,~ \lambda_2,~ \tan \beta$  and masses of all exotic particles. Other parameters and the VEV's are fixed in terms of these parameters. 
For our numerical study, we fix  $\tan \beta = 2.0,~ \lambda_2 = -0.1, ~\alpha_1 = 0.1, ~\alpha_2 = \alpha_3 = 0.1$
and perform a scan over range of masses as given in Table~\ref{tab:DMmassrange}.  
\begin{table}[htb]
\begin{center}
\begin{tabular}{c|c}
 \hline\hline
 Parameter&Range\\
 \hline \hline
 $M_{H_1^0} = M_{A_1}$&$(100-2000)$ GeV\\
 $M_{H_2^\pm} - M_{H_1^0}$ &$ (10-60)$ GeV\\
  $m_{q'} - M_{H_1^0}$ &$ (200-500)$ GeV\\
   $m_{n_\tau} - M_{H_1^0}$ &$ (0.001-1)$ GeV\\
   $\lambda_3$ & $(1.0-2.0)$\\
   $v^\prime$ & $(1.9-35)$ TeV \\
   $|\mu_3|$ & $(100-2000)$ GeV \\
 \hline
\end{tabular} 
\begin{tabular}{c|l}
 \hline
\multicolumn{2}{c}{$m_{n_e}$ and $m_{n_\mu}$}\\
 \hline \hline
 Case (i) (degenerate)&$m_{n_e}=m_{n_\mu}=m_{n_\tau}$ \\
  Case (ii) (small splitting) &$(m_{n_e}=m_{n_\mu})-m_{n_\tau} =$ Range (10 keV - 20 MeV) \\
    Case (iii) (large splitting) &$(m_{n_e}=m_{n_\mu})-m_{n_\tau} =$ Range (100 MeV - 10 GeV) \\
 \hline \hline
\end{tabular}
\end{center}
\caption{Range of masses considered for the scan in the scalar dark matter case, with $q'=d',~s',~b'$.  Three different cases of neutrino mass hierarchy are considered. Choosing $n_\tau$  as the 
lightest scotino  enhances signals in the collider studies, as discussed in Sec.~\ref{sec:collider}. }
\label{tab:DMmassrange}
\end{table}
We import this scan into {\tt MicrOmegas} to obtain the relic density and cross sections relevant to direct and indirect detection experiments. In a broader parameter range scan, the results of which are not discussed here, we found that only cases with $H_1^0$ and scotino being closely degenerate lead to the required relic abundance. This is  due to the role played by the co-annihilation channels in the freeze-out mechanism.   Furthermore, we take $n_\tau$ as the lightest scotino, as in the collider study $\tau$ channel has the advantage of having larger Yukawa coupling. We consider three different mass hierarchies between the scotinos; {\it i.e.} (i) $m_{n_e}=m_{n_\mu}=m_{n_\tau}$,  (ii) $(m_{n_\mu}, m_{n_e})-m_{n_\tau} \sim$ 10 keV - 20 MeV, and (iii) $(m_{n_\mu}, m_{n_e})-m_{n_\tau} \sim$ 100 MeV - 10 GeV, separately.  These three cases are identified, respectively, by greenish yellow, green and magenta dots in Fig.~\ref{fig:my_label} and \ref{fig:scalarID}.

In Fig. \ref{fig:my_label} we show the relic density and cross sections for direct detection (in a log plot) against $M_{H_1^0}$. On the left-hand side, the plot gives the relic density, with horizontal lines representing the Planck constraints within $1 \sigma$. The middle plot shows
the direct detection results,  compared to the results from the XENON \cite{XENON:2018voc} and PICO \cite{PhysRevD.100.022001,PhysRevLett.118.251301}, while in the right-handed plot we keep only the points that satisfy {\it both} direct detection and relic density constraints. The figure indicates that points corresponding to all Cases in Table \ref{tab:DMmassrange} satisfy both constraints, with the region of light and heavy $H_1^0$ being favoured by either degenerate, or small splitting among scotinos, while the region of intermediate $M_{H_1^0}$ is preferred by larger mass splittings among scotinos.  In all cases, we choose the scotino masses to be almost degenerate with scalar masses to enhance annihilations and yield correct relic density. For this analysis, we found the relic abundance to  be sensitive to variations in $\tan \beta = \frac{v_2}{v_L}$, while much less so  to $\lambda_1, \lambda_2, \alpha_1, \alpha_2$ and $\alpha_3$. For our analysis, we optimized the choice of $\tan \beta=2$ as it allows for lighter DM masses to satisfy the relic conditions.

In Fig. \ref{fig:scalarID} we plot the results of our analysis on indirect scalar dark matter detection, with the photon flux versus dark matter mass. All the points in the plots obey indirect detection bounds from Fermi-LAT \cite{Karwin:2016tsw}, while the black points satisfy relic, direct and indirect detection bounds together. This figure shows that, for all scotino mass splittings considered here, (i)--(iii), photon flux bounds from  indirect detection are consistent with a large region of the parameter space, indicating that a significant range of masses for both the $H_1^0/A_1$ dark matter candidates and the scotino splittings satisfy all experimental constraints.

We now turn to analyzing the parameter space available for the  fermionic candidates for dark matter, restricted by relic abundance, direct detection cross section, and indirect detection experiments.

\subsection{Scotino Dark Matter}
\label{subsec:scotinoDM}

An alternative to the above scalar dark matter scenario is to assume that one of the scotinos  is the lightest $R$-parity odd particle.  We shall chose  $n_\tau$ scotino as the dark matter candidate. This is because  it is more promising for collider studies (see Sec. \ref{sec:collider}). The  other two, $n_e$ and $n_\mu$  can be (i) degenerate with  $n_\tau$, or have  hierarchical mass. In the non-degenerate case, we shall consider two different cases  {\it i.e.} (ii) small mass splittings  with $(m_{n_\mu}, m_{n_e})-m_{n_\tau} \sim$ 10 keV - 20 MeV, or (iii) large mass splitting with $(m_{n_\mu}, m_{n_e})-m_{n_\tau} \sim$ 100 MeV - 10 GeV.  These three scenarios are indicated by the same colour-coded dots as in the scalar DM case reported in the previous subsection (Sec.~\ref{subsec:scalarDM}).  In this section, $R$-parity odd scalars, $H_1^0$ and $A_1$ are taken to be heavier with a mass splitting of around 100 GeV with $n_\tau$.  Smaller mass splittings are found to not yield the required relic density.   The  relevant annihilation channels are listed in Table~\ref{tab:scotino_annih}.  

\begin{table}[htb]
\begin{center}
\begin{tabular}{c}
 \hline\hline
         $n_\tau$ Annihilation Channels \\
        \hline
 $n_\tau {\bar n_\tau} \rightarrow \ell \bar \ell,~\nu\bar\nu,~q\bar q,~WW,~ZZ,~Zh,~ZH_1^0,~ZA_1^0,~hh$\\
 \hline\hline
 \end{tabular}
\end{center}
\caption{Annihilation channels for $n_\tau$. A specific channel will be available as long as the total mass of the final state is smaller than $2m_{n_\tau}$.
}
\label{tab:scotino_annih}
\end{table}
We perform a scan over the parameter region with the couplings, which do not affect the DM annihilation, fixed, and varying the masses of particles involved.  The values of the parameters used and the range of masses considered are given in Table~\ref{tab:scotinoMassrange}. 
\begin{table}[htb]
\begin{center}
\begin{tabular}{c|c}
 \hline\hline
 Parameter&Range (GeV)\\
 \hline \hline
 $m_{n_\tau}$&$(100-2000)$ \\
  $m_{q'} - m_{n_\tau}$ &$ (200-500)$\\
    $M_{H_2^\pm} - M_{H_1^0}$ &$ (10-70)$\\
    $v^\prime$ & $(1900-35000)$\\
 \hline
\end{tabular} \hspace*{5mm}
\begin{tabular}{c|l}
 \hline
\multicolumn{2}{c}{$m_{n_e}$ and $m_{n_\mu}$}\\
 \hline 
 Case (i) (degenerate)&$m_{n_e}=m_{n_\mu}=m_{n_\tau}$ \\
  Case (ii) (small splitting) &$(m_{n_e}=m_{n_\mu})-m_{n_\tau} =$ Range (10 keV - 20 MeV) \\
    Case (iii) (large splitting) &$(m_{n_e}=m_{n_\mu})-m_{n_\tau} =$ Range (100 MeV - 10 GeV) \\[1mm]
 \hline 
  \multicolumn{2}{c}{ }\\[-2mm]
 \multicolumn{2}{c}{$\tan \beta = 2.0,~ \lambda_2 = -0.1,~ \lambda_3 = 1.6, ~\alpha_1 = 0.1, ~\alpha_2 = \alpha_3 = 0.1$, $v_L = 5$ GeV, $\mu_3 = -320$ GeV}\\[2mm]
 \hline\hline
\end{tabular}
\end{center}
\caption{Range of masses considered for the scan in the scotino dark matter case, with $q'=d',~s',~b'$.  Three different cases of neutrino mass hierarchy are considered. $n_\tau$ is taken as the 
lightest, which would be important for the collider studies as discussed in Sec. \ref{sec:collider}. }
\label{tab:scotinoMassrange}
\end{table}

\begin{figure}[htb!]
    \centering
   \hspace{-0.5cm} \includegraphics[scale=0.36]{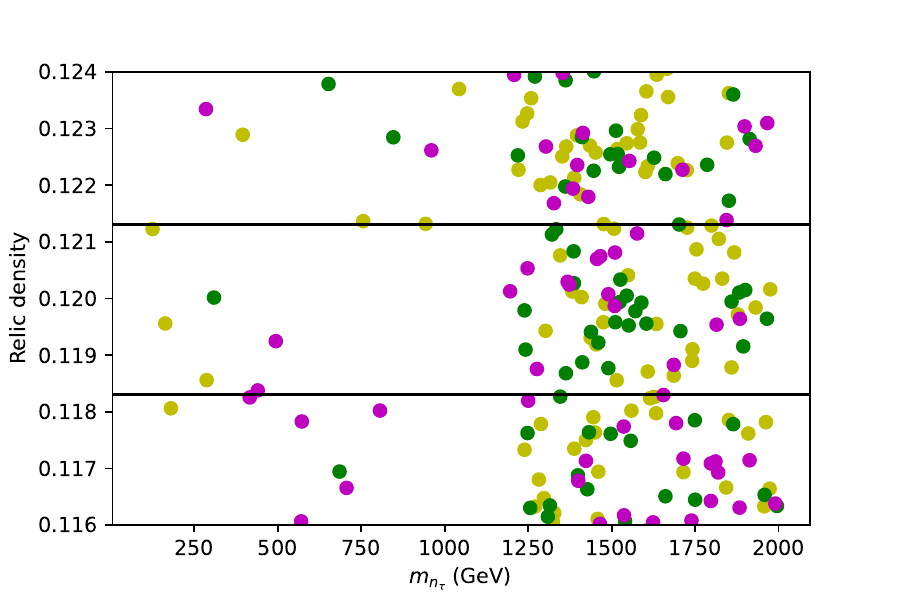}\hspace{-0.5cm}
    \includegraphics[scale=0.35]{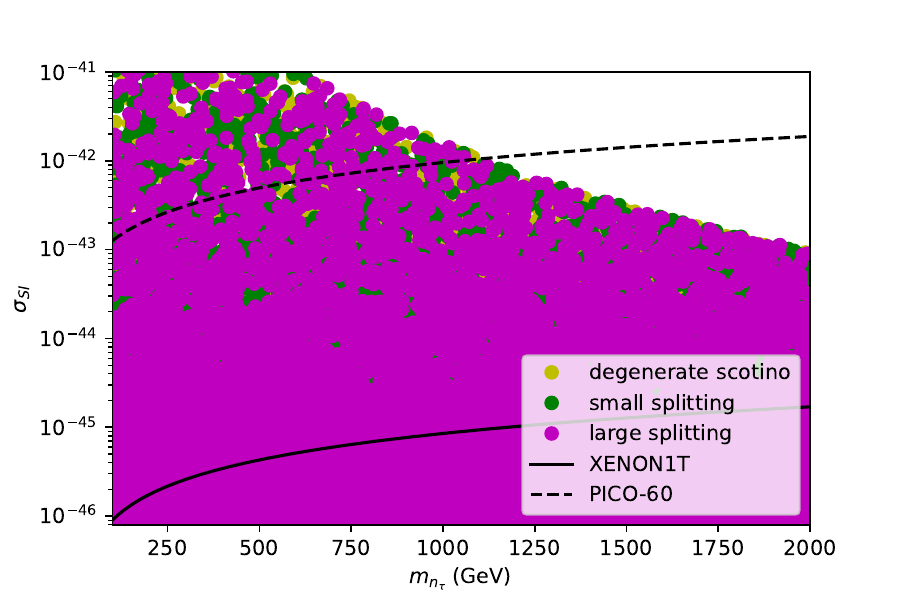}\hspace{-0.5cm}
    \includegraphics[scale=0.35]{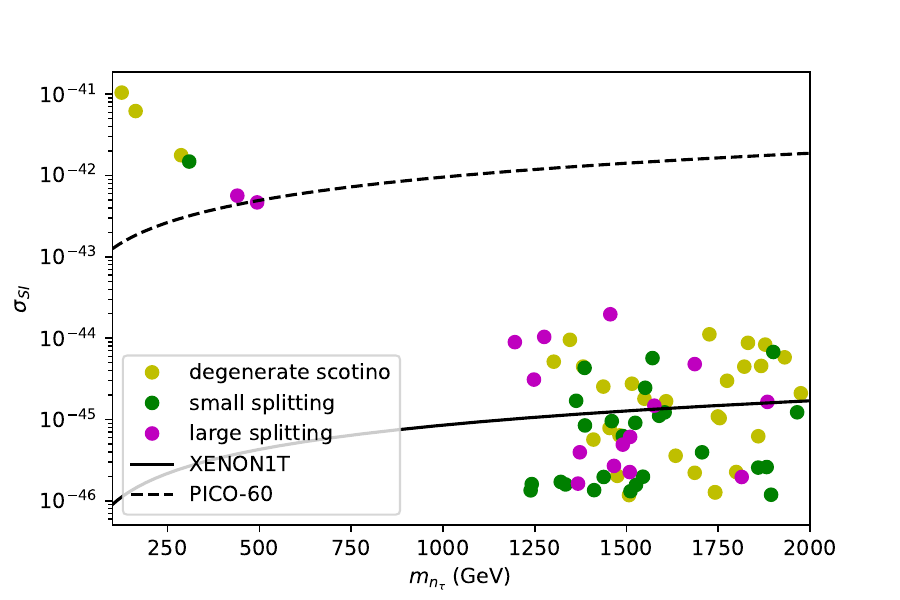}
 \caption{ (Left): Relic density vs. $m_{n_\tau}$ with horizontal black lines showing $1\sigma$ range $\Omega^{obs}_{DM}h^2$.
   (Middle): Direct detection cross section against $m_{n_\tau}$, along with bounds from XENON1T (in black solid) and PICO experiments (in black dashed).  
     (Right): Direct detection cross section against $m_{n_\tau}$, with points satisfied by $\Omega_{obs}$ within $1\sigma$.
Different colours correspond to different scotino mass hierarchy with Case (i) - degenerate scotino mass scenario (greenish yellow), Case (ii)- small mass splitting  (green) and Case (iii) - large mass splitting (magenta) given in Table~\ref{tab:DMmassrange}. }
    \label{fig:scotinoDM}
\end{figure}

In Fig. \ref{fig:scotinoDM} we plot the relic density and the  direct detection cross section, as a function of the scotino dark matter masses. On the plot at the left, we show the relic density, required to be within the horizontal lines representing $1 \sigma$ consistency with the measurements at the Planck experiment \cite{Planck:2018vyg}. The middle plot shows the parameter space restricted by the XENON \cite{XENON:2018voc} and PICO \cite{PhysRevD.100.022001,PhysRevLett.118.251301} experiments, while on the right, we indicate regions of the parameter space satisfying both direct detection and relic density constraints. As in the case of scalar dark matter, points with degenerate masses, small or larger mass splittings among the scotinos satisfy both constraints, but the masses are required to be $m_{n_\tau}>500$ GeV for PICO alone, while $m_{n_\tau}>1250$ GeV for the combined requirements of XENON1T and PICO.

\begin{figure}[htb!]
    \centering
    \hspace{-0.5cm}\includegraphics[scale=0.36]{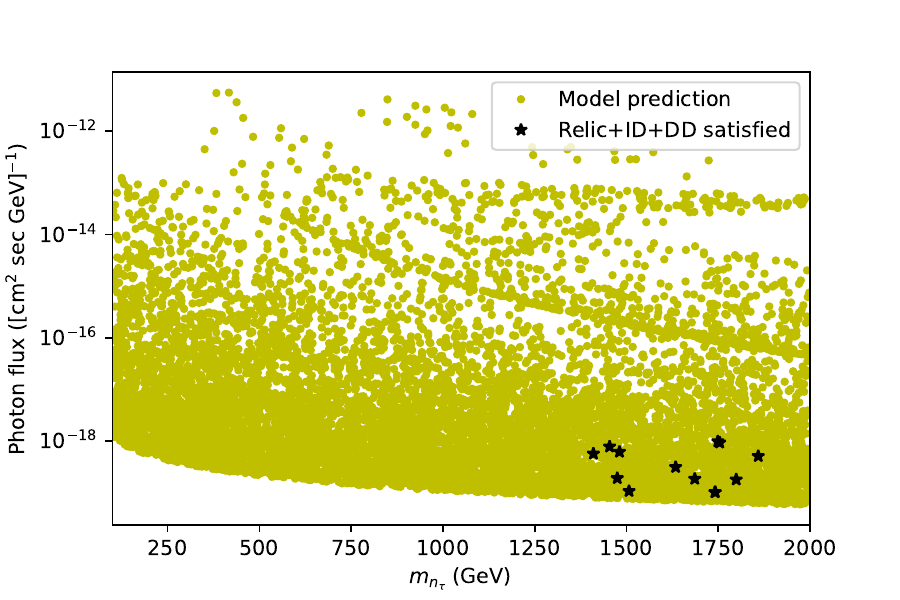}\hspace{-0.5cm}
    \includegraphics[scale=0.35]{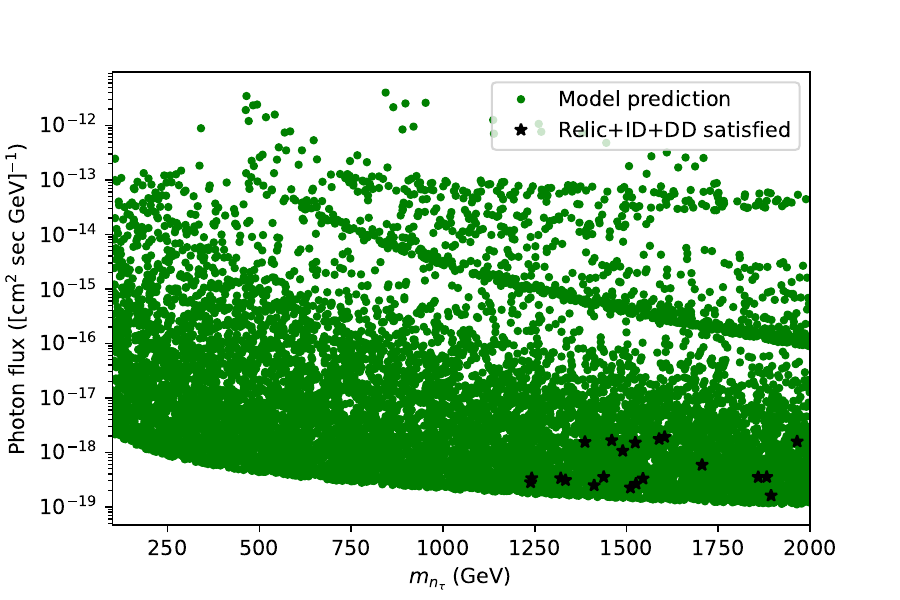}\hspace{-0.5cm}
    \includegraphics[scale=0.35]{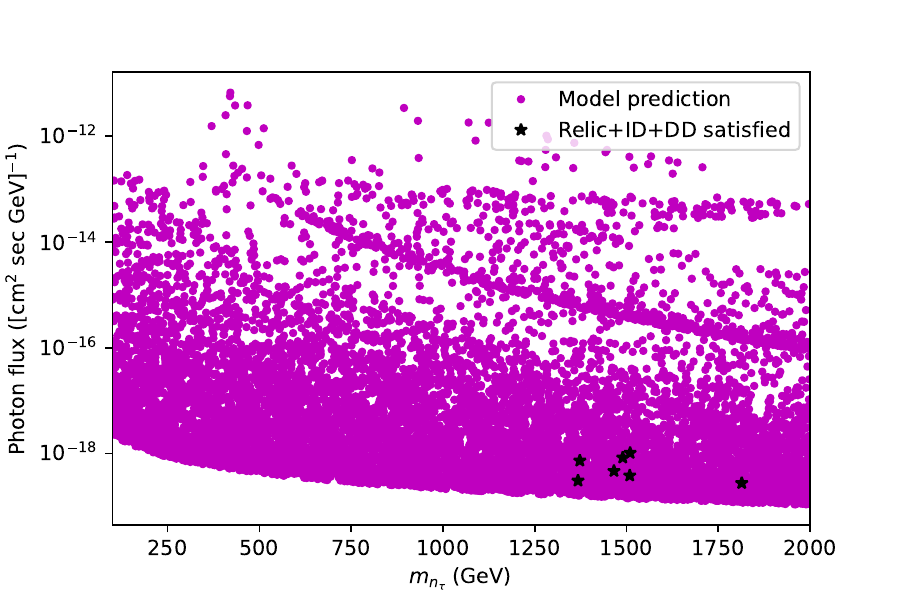}
    \caption{Indirect detection restrictions for scotino dark matter. The photon flux versus dark matter mass with Case (i) - degenerate scotino mass scenario (Left), Case (ii) - small mass splitting (Middle), and Case (iii) - large mass splitting (Right). All points satisfy constraints from Fermi-LAT \cite{Karwin:2016tsw}, while the black asterisks are also consistent with relic density + direct detection restrictions from PICO-60 and XENON1T experiments.}
    \label{fig:scotinoID}
\end{figure}

The relic is sensitive to $v^\prime = \sqrt{v_2^2 + v_R^2}$, as $Y_{n_\tau} = \frac{m_{n_\tau}}{v_R} \sim \frac{m_{n_\tau}}{\sqrt{v^{\prime 2}-v_2^2}}$, so we have varied both $v^\prime \in [1.9,35]$ TeV and $m_{n_\tau}$. This dependence can be understood as follows : for a fixed scotino DM mass,  if the $v^\prime$ increases within the above-mentioned range,  the Yukawa couplings decrease; thus the cross-section mediated by corresponding Higgs bosons would decrease and relic density increases. Moreover, for fixed $v^\prime$, if we increase scotino DM mass, the relic density for these associated channels decreases. But unlike the case of scalar dark matter,  the relic is insensitive to variations in $\tan \beta$ or to the scalar couplings.
To check compatibility with indirect detection experiments,  we plot the photon flux as a function of the scotino mass in Fig. \ref{fig:scotinoID}. Here again, the greenish yellow points (for degenerate scotino masses), green points (for small mass splittings among scotinos) and magenta points (for large scotino mass splittings)  represent regions of parameter space that satisfy indirect detection bounds, while the black asterisks are a subset of these points, which are consistent with relic density bounds and direct detection constraints.
%

\subsection{Degenerate Scalar-Scotino Dark Matter}
\label{subsec:degenerateDM}

The case where the scalar and scotino masses are the same represents the mixed/degenerate dark matter scenario. We have analyzed this case in detail and found it to be indistinguishable from the scalar dark matter scenario. The reason is that, for the case of scalar dark matter, the scotino masses had to be chosen to be almost degenerate with the DM scalar masses to yield the correct relic density. Thus the degenerate case introduces no new phenomenological features and we shall not investigate it further. 
\section{Discriminating amongst the different scenarios at the LHC }
\label{sec:collider}

We shall now analyze the signatures of the model at the LHC, through the direct production of the additional particles present beyond the SM within the two distinct dark matter scenarios presented above. The spectrum beyond the SM has two heavy charged scalars, two neutral pseudoscalars, and three neutral scalar particles plus the right-handed charged and neutral gauge bosons, $W_R$ and $Z^\prime$, as well as the exotic fermions, $d',~n$ and $\nu_R$. 

Exotic quark masses are constrained by experimental bounds. Searches for additional quarks have concentrated on vector-like states, as  additional hierarchical quark states conflict with the Higgs gluon fusion data. We briefly review the searches here.  Vector-like quarks  were searched for at ATLAS   \cite{ATLAS:2018ziw,ATLAS:2018tnt,ATLAS:2018uky,ATLAS:2018alq,ATLAS:2018mpo} and CMS \cite{CMS:2019eqb,CMS:2018wpl,CMS:2018zkf,CMS:2017ynm,CMS:2020ttz}, focusing mainly on the pair-production mode.  Mass limits were also obtained from differential cross sections measurements at the LHC \cite{ATLAS:2018ziw}. The most stringent limits at 95\% confidence
level (CL) on masses depend on the assumed branching ratio (BR) configuration: for 100\% BR for 
the decay $b' \to W t$ masses up to 1.35 TeV are excluded, while for 100\% BR for $b' \to Zb$ and $b' \to H b$, $b'$ masses up to 1.39 TeV and 1.5 TeV, respectively,  are excluded \cite{CMS:2019eqb}. The most recent study at ATLAS \cite{ATLAS:2022hnn} looks at pair production of $b'$ quarks in events with at least two electrons or
muons, where at least two same-flavour leptons with opposite-sign charges originate from the decay of the $Z$ boson, and where various branching ratios for the decay into a $Z, W$ or $H$ boson and a third-generation quark are considered. Mass limits of $m_{b'}> 1.42 $ TeV for singlets, and $m_{b'}> 1.20 $ TeV for doublets are obtained.

While these searches do not apply exactly to the exotic quarks in ALRM (which are part of doublets, but with exotic quantum numbers),  for our benchmarks we set conservative mass limits on the exotic fermions, $m_{q'}>1.45$ TeV, $q'=d', s', b'$. The masses of scalar particles are given in Sec.~\ref{subsec:scalarDM}. The masses of fermions  can be set independently from the other sectors. In principle, all these particles can be produced at the LHC in pairs, in association with each other, or with other SM particles, as long as $R$-parity is conserved.

As the model depends on several parameters, we choose two benchmarks corresponding to the scalar dark matter case,  {\bf BP1} and  {\bf BP2}, and two for the scotino dark matter,  {\bf BP3} and  {\bf BP4}, to showcase our results. These benchmarks are parameter points which satisfy dark matter constraints, are consistent with the analysis in the previous subsections \ref{subsec:scalarDM} and \ref{subsec:scotinoDM}, and show some promise toward being observed at the LHC. For this,  the masses of the new particles are set such that the DM candidates (the lightest scalar and/or pseudo-scalar and the scotinos) are in the ${\cal O} \sim 1$ TeV range. Note that, while the Higgs masses could be in the sub-TeV range, correct relic density is obtained only for the case where they are almost degenerate in mass with the scotinos. As dark matter constraints require the scotino masses to be $m_{n_\tau}>1250$ GeV, this pushes the Higgs masses in the TeV region as well. Similarly, the lightest charged Higgs bosons and the exotic down-type quarks are also chosen to have masses in the ${\cal O} \sim 1$ TeV  range. This enhances their production, followed then by the cascade decay down  to the DM plus other SM particles.  Note that for the scotino cases, we take the tau scotino as the LSP to enhance production rates due to the larger Higgs-$\tau$ Yukawa coupling. (Note that in {\bf BP1} and  {\bf BP2} the scotino and scalar masses are almost degenerate, to enhance co-annihilation as discussed in Sec.~\ref{subsec:scalarDM}.)  

The presence of charged Higgs bosons and exotic quarks in the low-TeV  range of mass presents the  possibility that their presence can be explored at the LHC.  On the other hand the new gauge bosons $W_R$ are expected to be a few TeV in mass so as to respect the existing LHC limits on $Z^\prime$ dilepton decays \cite{ATLAS:2017fih}. This bound is satisfied by all the points we have considered in the DM analysis.   
The values for the model parameters corresponding to benchmark points {\bf BP1},  {\bf BP2}, {\bf BP3} and  {\bf BP4}, satisfying all the DM constraints from the study in Sec.~\ref{subsec:scotinoDM} are given in Table~\ref{tab:BPs}.
\begin{table}[htb]
    \centering
    \begin{tabular}{c|c|c|c|c|c|c|c}
    \hline\hline
   BP's&  $\tan \beta $&$ v_R ${ (TeV)}&$\lambda_2 $&$\lambda_3$&$ \alpha_1$&$  \alpha_2 = \alpha_3 $&$ \mu_3 $
   \\
   \hline
    {\bf BP1}&2&12.9 & -0.1&1.2& 0.1&0.1& -100\\
   {\bf BP2}&3&14.4 & -0.1&1.4& 0.01&0.1& -340\\
   {\bf BP3}&2&14 & -0.001&1.6& 0.1&0.1& -320\\
   {\bf BP4}&10&13 & -0.01&1.6& 0.1&0.1& -2800\\
         \hline
         \end{tabular}
\begin{tabular}{c|c|c|c|c|c|c|c}
    \hline
 &\multicolumn{7}{c} {Masses in GeV}\\ 
 \hline
   BP's&  $M_{H^0_1} = M_{A^0_1}$ & $m_{n_\tau}$ &$m_{n_\mu}$ & $m_{n_e}$ &$m_{d'}$ &$m_{s'}$ &$m_{b’}$   \\
   \hline
    {\bf BP1}&1230&1230.004&1230.085&1230.453&1465&1704&1712 \\
   {\bf BP2}&1428&1428.039&1428.481&1428.868&1764&1877&1902 \\
   {\bf BP3}&1258.51&1210&1215&1220&1600&1700&1800\\
   {\bf BP4}&1603.97&1395&1399&1402&1600&1700&1800\\
         \hline\hline
         \end{tabular}
\caption{Benchmark points compatible with the scalar ({\bf BP1} and {\bf BP2}) and scotino ({\bf BP3} and {\bf BP4}) DM constraints, selected for the study of collider effects. }
         \label{tab:BPs}
         \end{table}
With these choices for the parameters, the  masses of right-handed gauge bosons and the other heavy scalars corresponding to the above {\bf BP}'s are listed in Table~\ref{tab:scalarmass}.
\begin{table}[htb]
    \centering
    \begin{tabular}{c|c|c|c|c|c|c|c|c}
    \hline
    &\multicolumn{7}{c|}{Masses in GeV}&DM Relic\\ 
    \cline{2-8}
   BP's& $M_{H^\pm_1}$  & $M_{H^\pm_2}$ & $M_{A^0_2}$& $M_{H^0_2}$  & $M_{H^0_3}$ & $M_{W^\pm_R}$ & $M_{Z^\prime}$&$\Omega h^2$\\
         \hline
   {\bf BP1}&    1515.839& 1277&1515.882&1517.035&20137.437&4212& 5054.39&0.119\\
  {\bf BP2}&   1520.740&1471&1520.912&1520.042&21750.899&4687.23&5624.63&0.121\\
  {\bf BP3}&   2814.18&1258.67&2814.24 &19799.5&2814.72&4571.56&5485.88& 0.120\\
      {\bf BP4}& 16123.30&1604.60&16123.30& 23255.30& 16123.10& 4245.22&5094.27& 0.119\\
         \hline
    \end{tabular}
    \caption{Masses of the gauge bosons and heavy scalars corresponding to the four Benchmark Points considered in this study.}
    \label{tab:scalarmass}
\end{table}
The choice for the masses of new particles such that  the DM candidates (the lightest scalar and/or pseudo-scalar and the scotinos) in the low TeV range, the lightest charged Higgs   also arranged to also have masses in the same range, facilitates the production of these latter degrees of freedom, which then decay  into the DM and other SM particles. 

For production mechanisms, the only available processes are the pair productions of $q'\bar q'$ (with $q'=d'~,s',~b'$), $H_2^+H_2^-$ and $W_R^+W_R^-$ at the LHC with centre of mass energy 14 TeV  and high luminosity, as these particles have negative $R$-parity, necessary for decay into the lightest $R$-parity negative particle. The corresponding cross sections  are listed in Table \ref{tab:crosssection}.  For the scalar DM, the dominant cross section comes from $pp \rightarrow \bar{q^\prime} q^\prime $, while for the scotino DM case, the only viable process is the charged Higgs pair production, since the gauge bosons $W_R$ are heavy\footnote{The same is true for the production cross section $pp \to W_R^\pm H_2^\mp$ which is much smaller than that of $pp \to H_2^+H_2^-$. Further, we do not include processes like $pp\to q'q'$ with $q'\to H_2^-u\to \tau n_\tau u$, considering its large jet multiplicity. }. In Table \ref{tab:crosssection}, the dash $-$ means that the cross section is not important for the {\bf BP} under consideration. 
\begin{table}[htb]
    \centering
    \begin{tabular}{c|c|c|c|c}
    \hline\hline
    &\multicolumn{4}{c}{Cross section (fb)}\\
    \hline
       Processes &{\bf BP1}& {\bf BP2}&{\bf BP3}&{\bf BP4}\\
       \hline
       $pp \rightarrow \bar{d^\prime} d^\prime $ &32.1 &6.3 & $-$ & $-$\\
              $pp \rightarrow \bar{s^\prime} s^\prime $ & 8.7 & 3.5 & $-$ & $-$\\
                     $pp \rightarrow \bar{b^\prime} b^\prime $ & 8.3 & 3.1 & $-$ & $-$\\
       \hline
       $pp \rightarrow H_2^+ H_2^-$ & $-$ & $-$ & $8.341$ &$0.7918 $\\
       \hline
       $pp \rightarrow W_R^+ W_R^- $ & $-$ & $-$ & $9.64\times 10^{-7}$& $4.96\times 10^{-8}$\\
       \hline       \hline
           \end{tabular}
    \caption{Cross sections of the pair production of exotic quarks, lightest charged Higgs  and the right-handed charged gauge bosons at the 14 TeV LHC for the four different {\bf BP}'s considered.}
    \label{tab:crosssection}
\end{table}
The relevant decay branching fractions of $q'$ and $H^+_2$ are given in Table~\ref{tab:BRs}.
\begin{table}[htb]
    \centering
    \begin{tabular}{c|c|c|c|c|c}
    \hline
    \hline
    \multicolumn{3}{c|}{Scalar DM}&    \multicolumn{3}{c}{Scotino DM}\\ \hline
       Channel &\multicolumn{2}{c|}{BR (\%)}& Channel&\multicolumn{2}{c}{BR (\%)}\\ \cline{2-3}\cline{5-6}
       &{\bf BP1}&{\bf BP2}&&{\bf BP3}&{\bf BP4}\\
       \hline
       $ d^\prime\to b~H_1^0/A_1^0 $ &70.7 & 65.1 &   $ H_2^+\to \tau~n_\tau $ & 96.0 & 87.7 \\
      $ s^\prime\to b~H_1^0/A_1^0 $ &56.3& 55.9  & $ H_2^+\to \mu~n_\mu $ & 2.1 & 7.4 \\
      $ b^\prime\to b~H_1^0/A_1 $ & 55.7 & 55.2 &$ H_2^+\to e~n_e $ & 1.9 & 4.9 \\
       \hline       \hline
           \end{tabular}
    \caption{Branching fractions of the exotic quark decay to $b$ quark, and the charged scalar to the leptonic channels, relevant for the four different {\bf BP}'s considered.}
    \label{tab:BRs}
\end{table}

For the case of  scalar DM we consider benchmarks {\bf BP1} and {\bf BP2} and study $pp \to q'q'$,    with $q'\to b~H_1^0/A_1$. The light quark decay channels are negligible compared to the $b$ decay channel. The final state is thus $b\bar b ~DM~DM$, with $DM=(H_1^0/A_1)$ not detected, and we consider the SM background $pp\to b\bar b~\nu\bar\nu$, where we sum over the neutrino flavours ($\nu=\nu_e+\nu_\mu+\nu_\tau$) and assume $b$-tagging. The differences in cross sections between {\bf BP1} and {\bf BP2} are  due to  the differences in the masses of the exotic quarks, assumed heavier in  {\bf BP2} than in {\bf BP1}.


For the case of scotino dark matter,  we  consider the production chain $pp\to H_2^+H_2^-$, followed by the decay  $H_2^\pm \to n_\tau \tau^\pm$ (with the $\tau$ scotino as the lightest to enhance production rates due to the larger Higgs-$\tau$ Yukawa coupling), leading to the final state process $pp\to H_2^+H_2^-\to n_\tau n_\tau \tau^+\tau^-$.  The decay rates for $H_2^\pm \to n_\tau \tau^\pm$ for {\bf BP3} and {\bf BP4} are included in Table \ref{tab:BRs}. We limit our feasibility study to the parton level, and consider the SM background $pp\to \nu \bar{\nu}\tau^+\tau^-$. 
The corresponding cross sections for the signal and background for {\bf BP1},  {\bf BP2}, {\bf BP3} and {\bf BP4} are listed in Table~\ref{tab:crosssection3}.
\begin{table}[h]
    \centering
    \begin{tabular}{c|c|c|c|c}
    \hline\hline
    &\multicolumn{4}{c}{Cross section (fb)}\\
    \hline
       Processes &{\bf BP1}&{\bf BP2}&{\bf BP3}&{\bf BP4}\\[.8mm]
       \hline\hline
       $pp \rightarrow q'q'\to b\bar b~(H_1^0/A_1)~(H_1^0/A_1)$ & $20.3 $ & 3.7&$-$&$-$  \\
       \hline
       $pp \rightarrow \bar{\nu} \nu b {\bar b}$ & \multicolumn{4}{c}{204630}  \\
       \hline 
       $pp \rightarrow H_2^+ H_2^-\to n_\tau n_\tau \tau\tau$&$-$&$-$ & $5.52 $ & $0.58$  \\
       \hline
       $pp \rightarrow \bar{\nu} \nu \tau\tau$ & \multicolumn{4}{c}{506.11} \\
       \hline       \hline
           \end{tabular}
            \caption{Cross section times BR for the selected channel and the corresponding SM background  at the 14 TeV LHC for the four chosen {\bf BP}'s. }
    \label{tab:crosssection3}
\end{table}

For the collider analysis, we used {\tt MadGraph5} \cite{Alwall:2014hca} and {\tt MadAnalysis5} \cite{Conte:2012fm} to generate events with the input parameters corresponding to the {\bf BP}'s considered. We use the parameters as defined in the software, before imposing our cuts to enhance the signal over the background.  

\begin{figure}[htb!]
\centering
\includegraphics[scale=0.45]{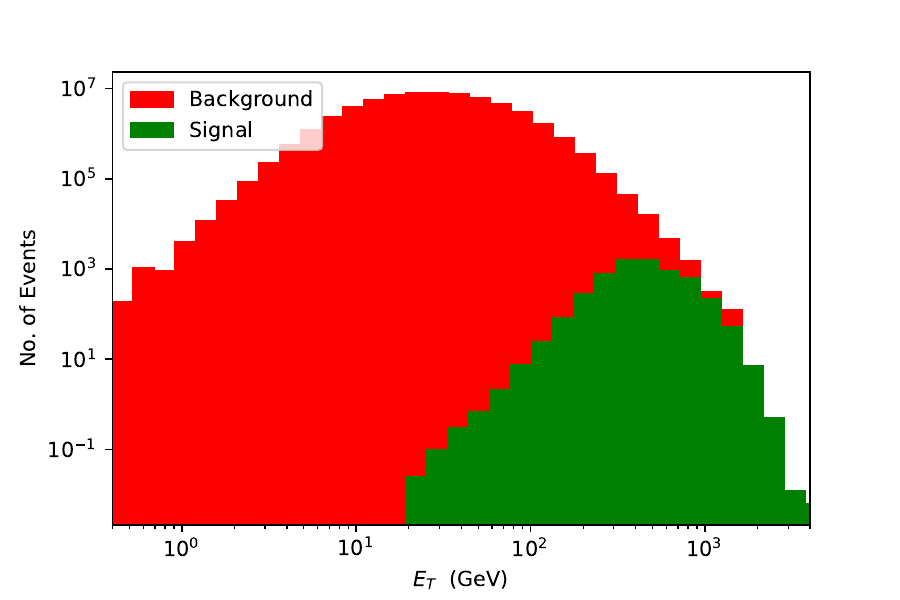}\qquad
\includegraphics[scale=0.45]{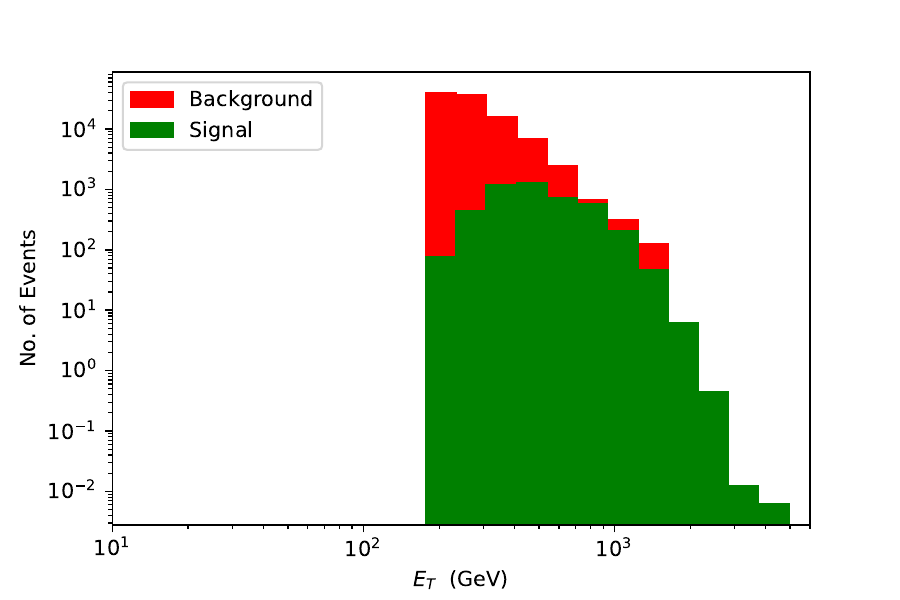} \\
\includegraphics[scale=0.45]{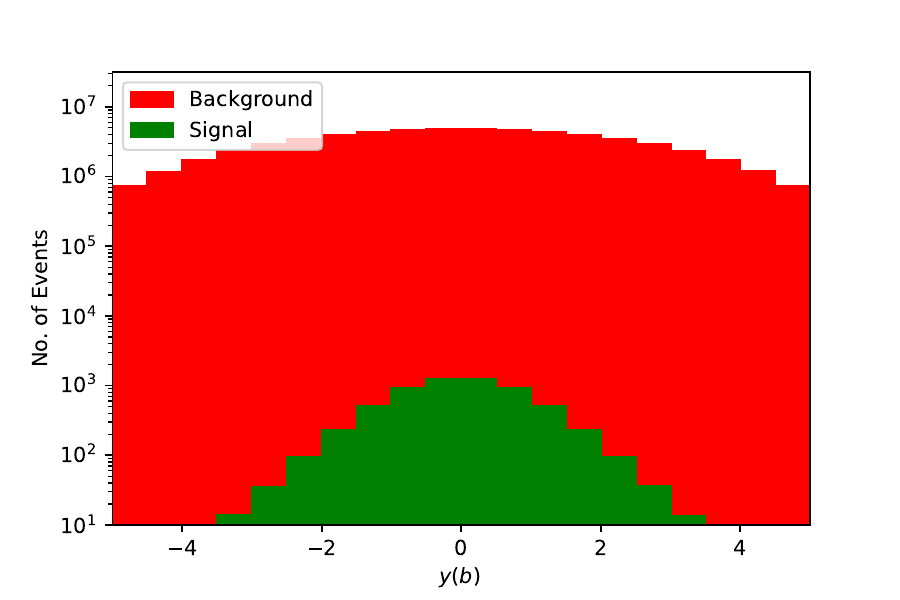}\qquad
\includegraphics[scale=0.45]{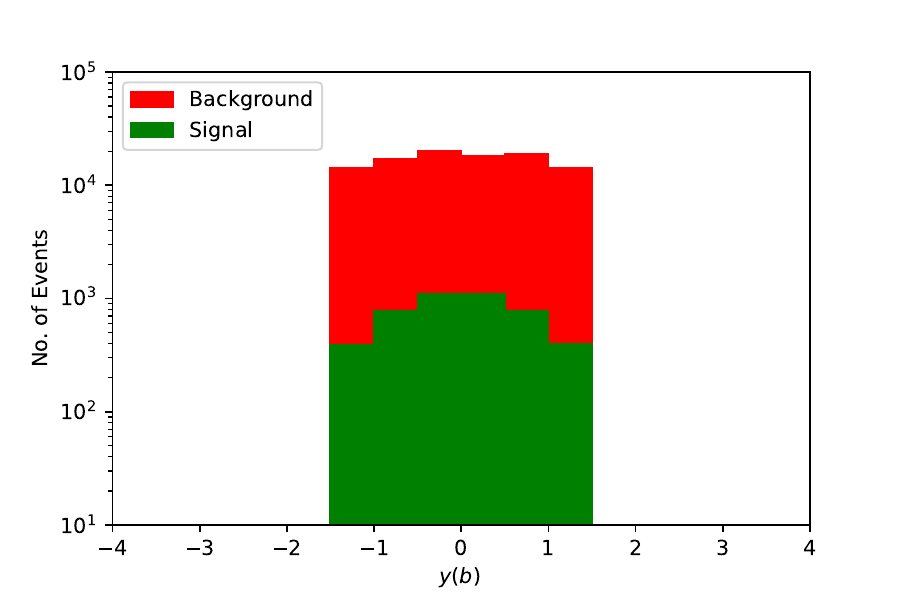}\qquad
\includegraphics[scale=0.45]{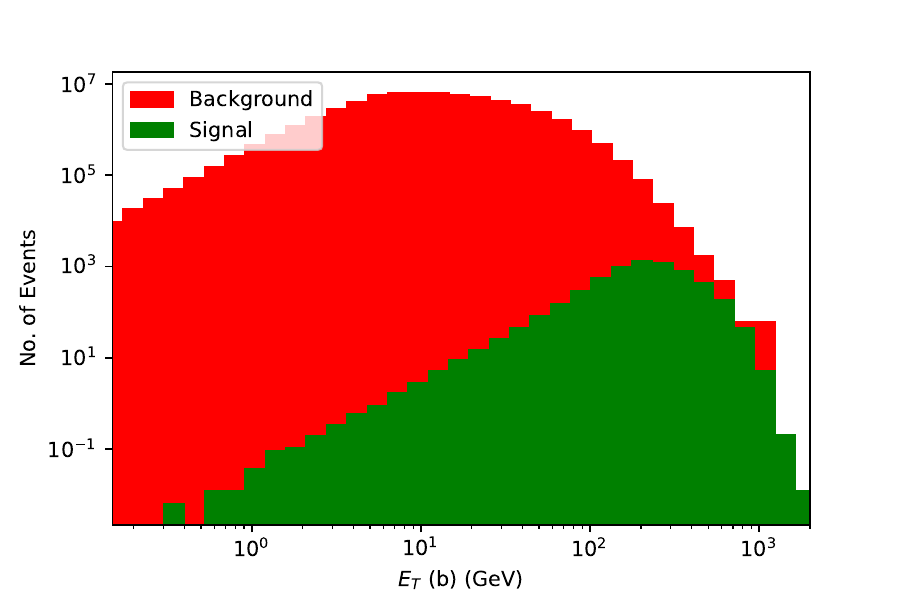}\qquad
\includegraphics[scale=0.45]{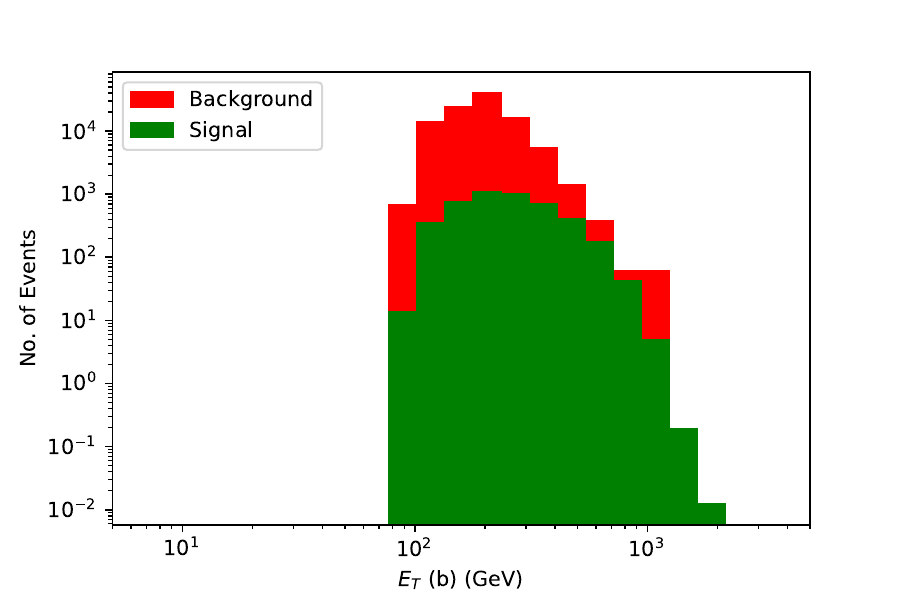}\qquad\\
\includegraphics[scale=0.45]{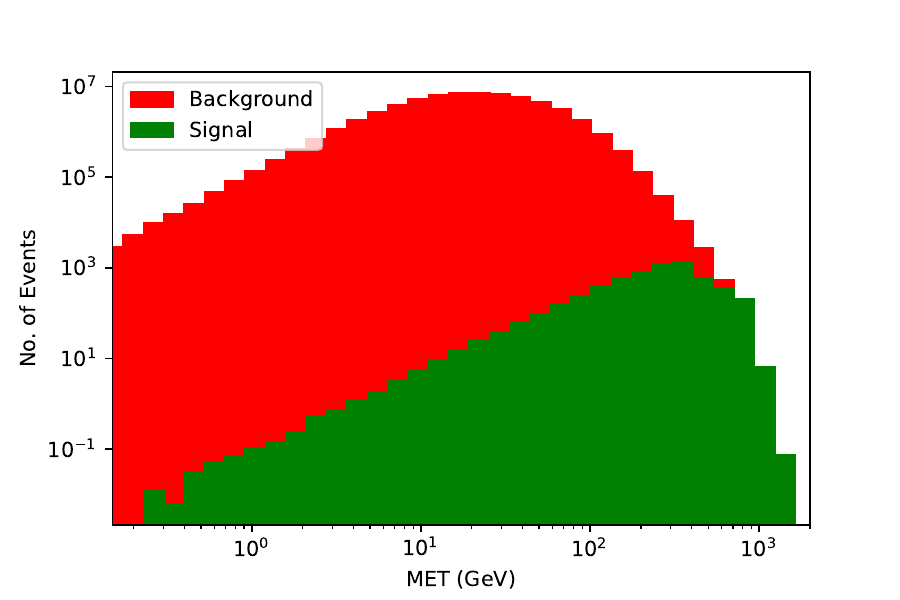}
\includegraphics[scale=0.45]{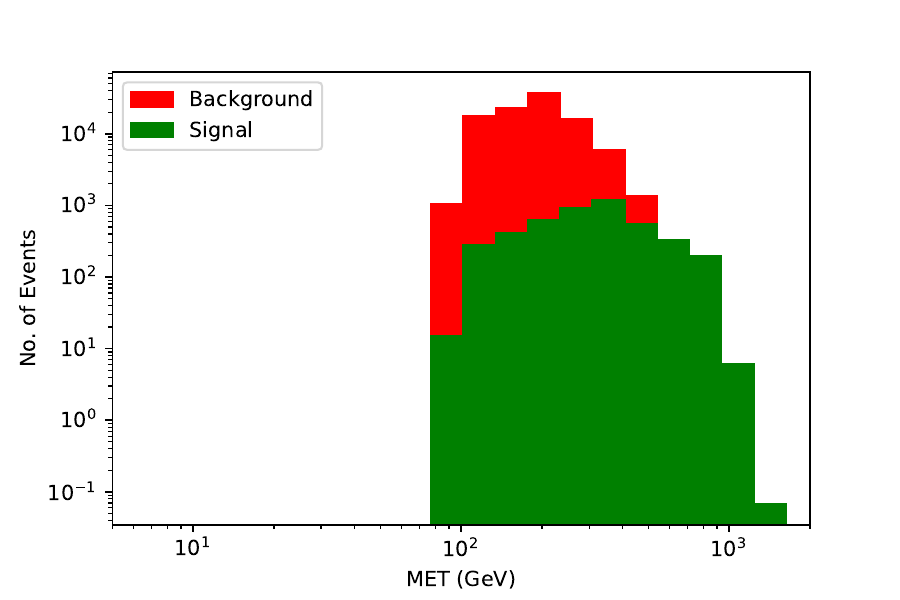}
\vskip-0.5cm
\caption{Parton level event distributions for $pp\to \bar{q^\prime} q^\prime \to \bar{b} b H_1^0 H_1^0 / A_1^0 A_1^0$ for {\bf BP1} normalized to an integrated luminosity of 300 fb$^{-1}$ along with the SM background (in red), here $q^\prime$ represents all the exotic quarks i.e., $d^\prime, s^\prime, b^\prime$.  
(Left column, top to bottom): the total transverse energy $E_T$, the rapidity $y$, the total transverse energy of the $b$ quark, $E_T(b)$, and the missing transverse energy $\slashed{E}_T$, before cuts.
(Right column, top to bottom): the same distributions after implementing the cuts in Table \ref{tab:SB_BP1}, which reduce the background. }
\label{fig:scalar_BP1}
\end{figure}

\begin{figure}[htb!]
\centering
\includegraphics[scale=0.45]{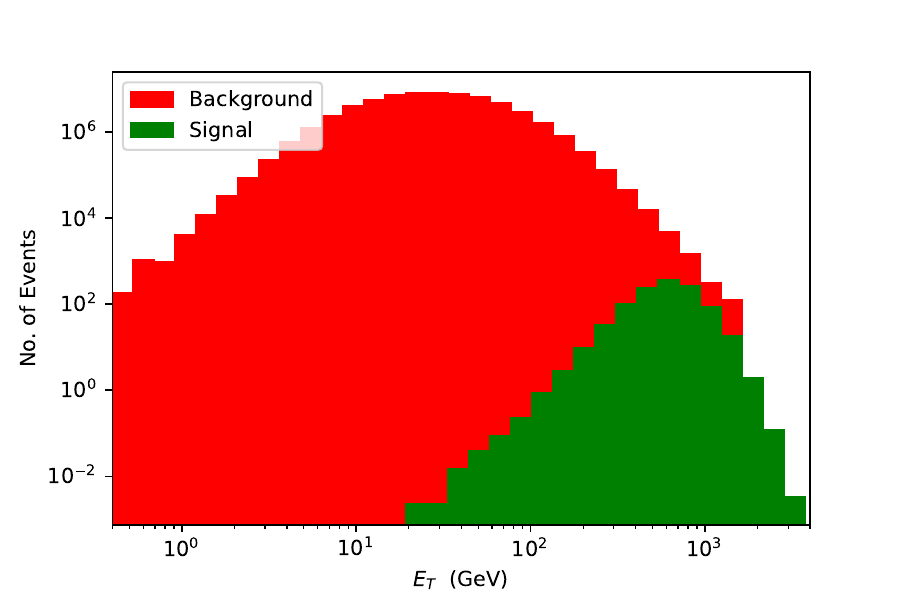}\qquad
\includegraphics[scale=0.45]{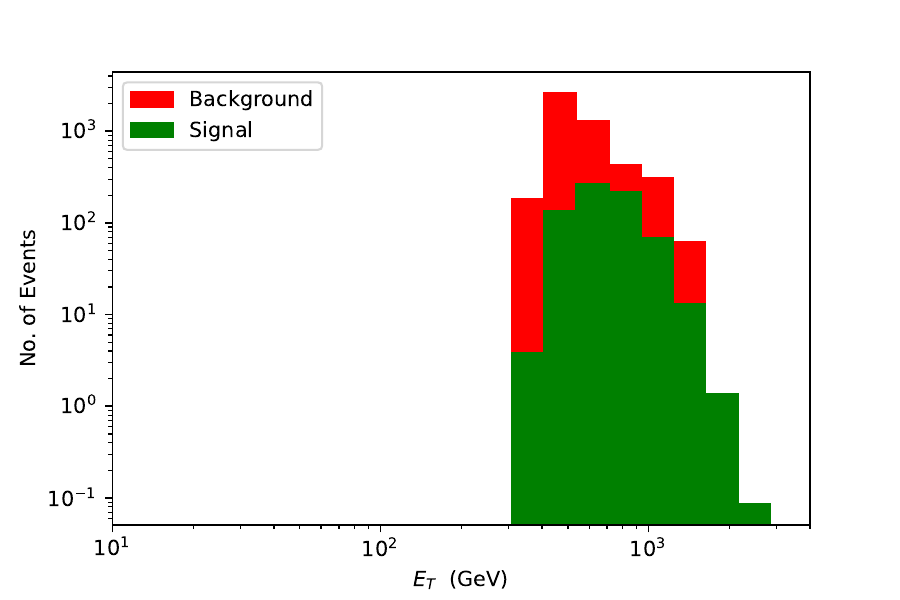}\qquad \\
\includegraphics[scale=0.45]{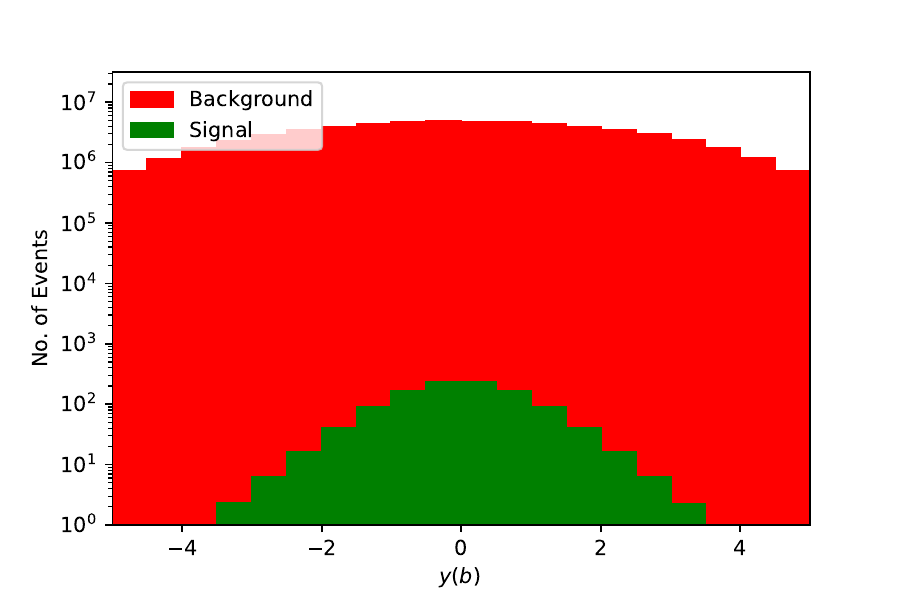}\qquad
\includegraphics[scale=0.45]{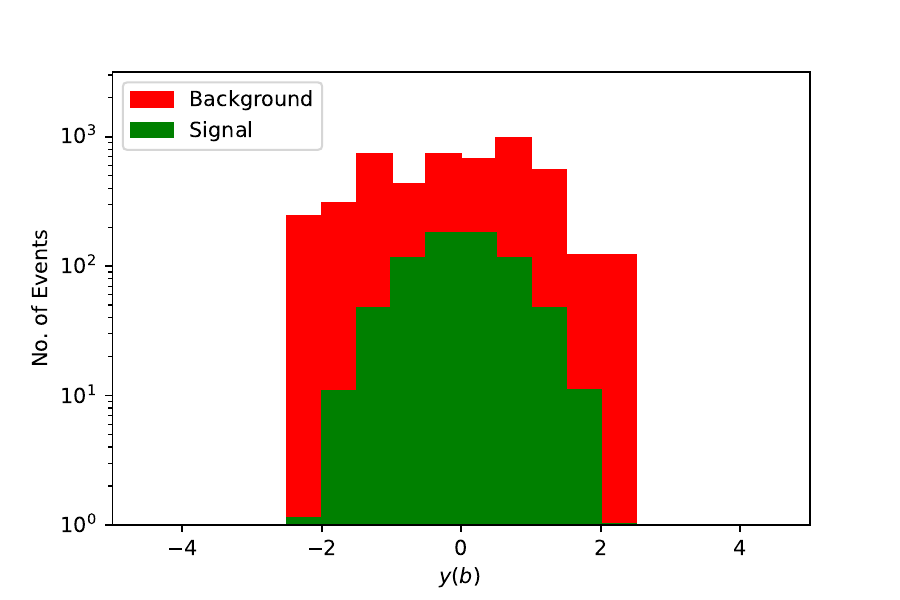}\qquad
\includegraphics[scale=0.45]{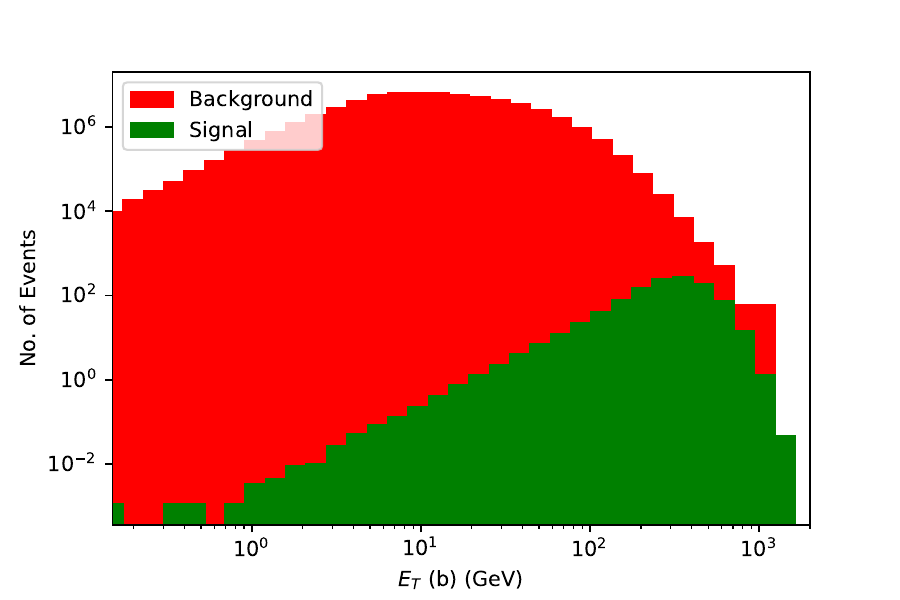}\qquad
\includegraphics[scale=0.45]{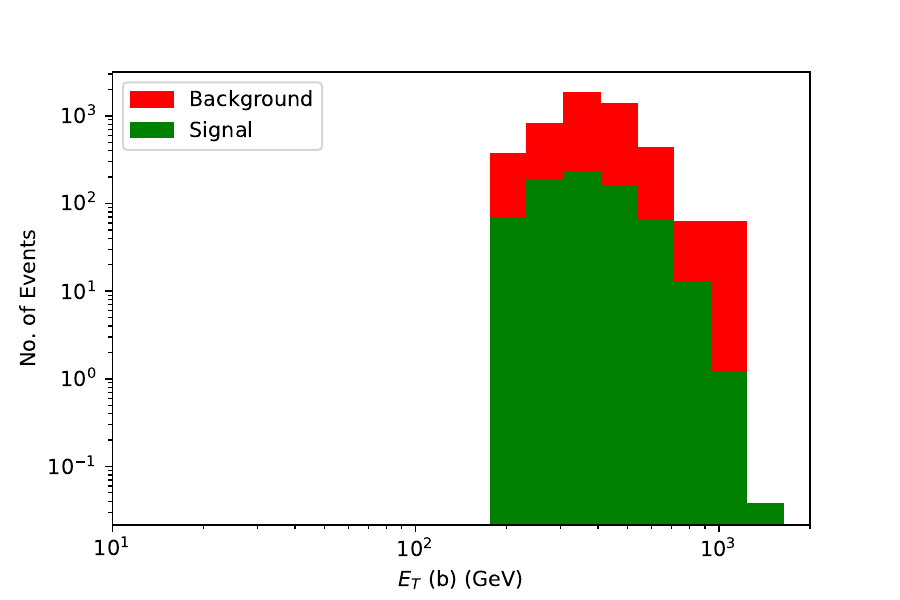}\qquad\\
\includegraphics[scale=0.45]{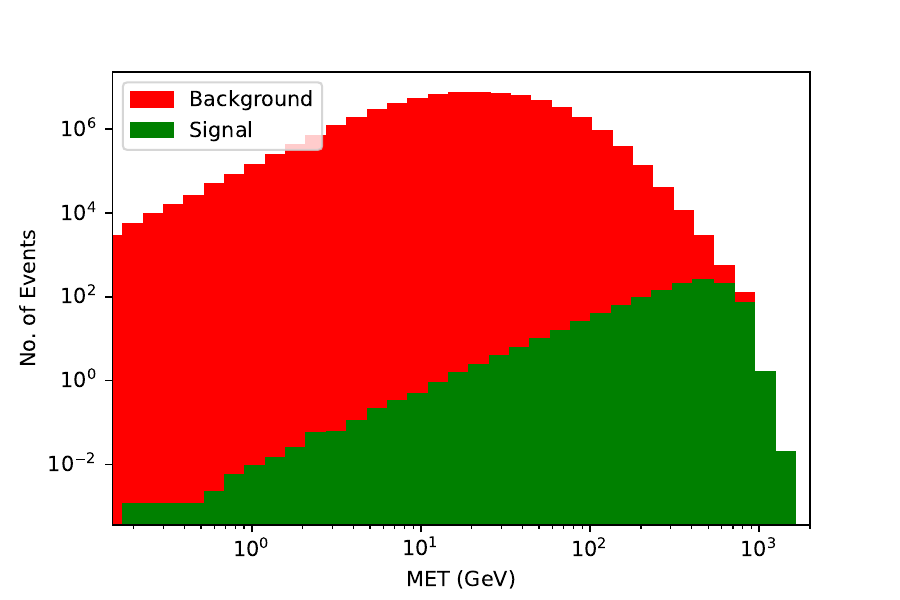}
\includegraphics[scale=0.45]{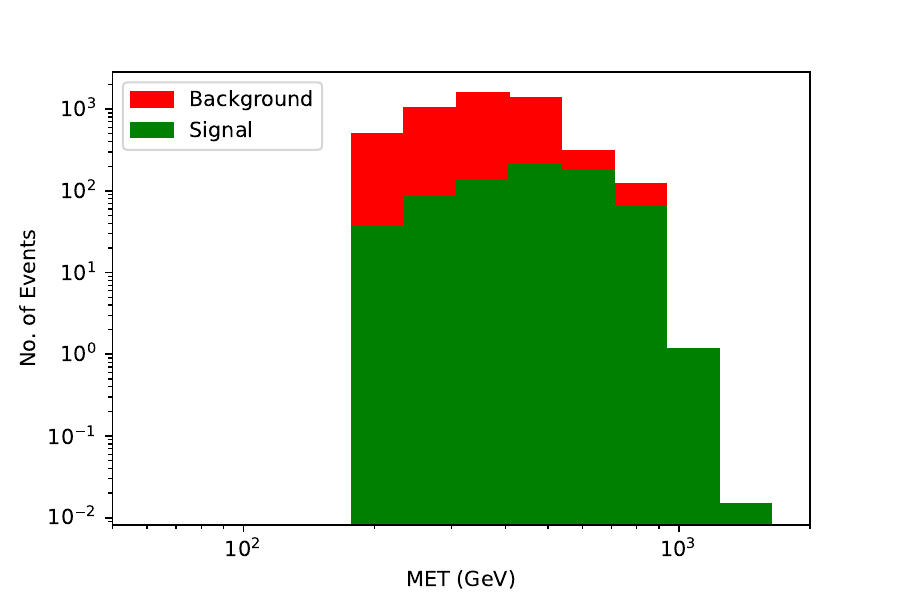}
\vskip-0.5cm
\caption{Parton level event distributions for $pp\to \bar{q^\prime} q^\prime \to \bar{b} b H_1^0 H_1^0 / A_1^0 A_1^0$ for {\bf BP2} normalized to an integrated luminosity of 300 fb$^{-1}$ along with the SM background (in red), here $q^\prime$ represents all the exotic quarks i.e., $d^\prime, s^\prime, b^\prime$.  
(Left column, top to bottom): the total transverse energy $E_T$, the rapidity $y$, the total transverse energy of the $b$ quark, $E_T(b)$, and the missing transverse energy $\slashed{E}_T$, before cuts.
(Right column, top to bottom): the same distributions after implementing the cuts in Table \ref{tab:SB_BP2}, which reduce the background.  }
\label{fig:scalar_BP2}
\end{figure}

First  we analyze the scalar dark matter case. Considering a cut and count method to improve the signal significance, in Table~\ref{tab:SB_BP1} and \ref{tab:SB_BP2} we give the number of signal and background events and the corresponding significance at each stage of selection, for an integrated luminosity of 300 fb$^{-1}$. 
The significance is defined as
\begin{equation}
S=\frac{N_{\rm signal}}{\sqrt{N_{\rm signal}+N_{\rm background}}},
\label{eq:significance}
\end{equation}
and it yields a sufficiently large significance even at the early stages of HL-HLC. However, this analysis does not include any detector effects, and possible additions to background from processes like $pp\to bb, ~bbjj, ~ jj, ~ jjjj$, which can mimic the final state of $bb+\slashed{E}_T$ when parton showering, jet formation and detector efficiencies are considered. We do not attempt a full analysis including these effects here. These effects could reduce the significance quoted in Table~\ref{tab:SB_BP1}  and \ref{tab:SB_BP2}. For example, an assumed total event reconstruction efficiency of 30\% and an increase in the background events by about 50\% would change the final significance from 13.8 to 3.4 at the integrated luminosity of 300 fb$^{-1}$, in the case of {\bf BP1}.
\begin{table}[htb!]
    \centering
    \begin{tabular}{c|c|c|c}
    \hline\hline
       Selection   criteria      &\multicolumn{2}{c|}{Fiducial Cross section (fb)}&{Significance $S$}\\ \cline{2-3}
   (BP1)  & {Signal} & {Background}&{ @ 300 fb$^{-1}$} \\ \cline{4-4}
       \hline\hline
       Initial (no cut) & 20.3&2046.6&7.7 \\
reject : $E_T< 200$ GeV  &19.6&1175.6&9.8\\
       reject : $E_T( b)< 100$ GeV  &18.1&777.5&11.1\\
       reject : $\slashed{E}_T< 100$ GeV  &16.2&464.6&12.8\\     
       reject : $-1.5<y(b)< 1.5$  &15.1&341.4&13.8\\
       \hline
    \end{tabular}   
     \caption{Fiducial cross section of the signal ($pp\to d'd'\to bb~H_1^0H_1^0/A_1A_1$) and the SM background ($pp\to bb\nu\nu$) corresponding to {\bf BP1}  at 14 TeV LHC, with and without the selection criteria considered, and the signal significance at the HL-LHC.}
    \label{tab:SB_BP1}
\end{table}
In Fig.~\ref{fig:scalar_BP1}  we show the relevant kinematic distributions for {\bf BP1}. The first column refers to the case before the selection criteria are applied, while second column corresponds to the signal and background events, after the listed selection cuts  are applied. 
We repeat the same analysis for {\bf BP2}. The cuts implemented are given in Table \ref{tab:SB_BP2} and the plots corresponding to signal and background events before and after cuts are shown in Fig. \ref{fig:scalar_BP2}.
\begin{table}[htb!]
    \centering
    \begin{tabular}{c|c|c|c}
    \hline\hline
       Selection   criteria      &\multicolumn{2}{c|}{Fiducial Cross section (fb)}&{Significance $S$}\\ \cline{2-3}
   (BP1)  & {Signal} & {Background}&{ @ 300 fb$^{-1}$} \\ \cline{4-4}
       \hline\hline
       Initial (no cut) & 3.7&2046.6&1.4 \\
reject : $E_T< 200$ GeV  &3.3&79.4&6.3\\
       reject : $\slashed{E}_T< 200$ GeV  &2.8&29.5&8.4\\
       reject : $E_T(b)< 200$ GeV  &2.4&16.4&9.4\\   
              \hline
    \end{tabular}    \caption{Fiducial cross section of the signal ($pp\to d'd'\to bb~H_1^0H_1^0/A_1A_1$) and the SM background ($pp\to bb\nu\nu$) corresponding to {\bf BP2}  at 14 TeV LHC, with and without the selection criteria considered, and the signal significance at the HL-LHC.}
    \label{tab:SB_BP2}
\end{table}

The above analysis illustrates that HL-LHC can potentially discover the presence of exotic quarks that appear in the ALRM version considered in this study, even at 300 fb$^{-1}$. 

Turning to the scotino DM case, we analyze the signal and background.  Defining the signal significance as in Eq.~\ref{eq:significance},
the fiducial cross sections and the signal significance corresponding to the integrated luminosities of 300 fb$^{-1}$ and 3000 fb$^{-1}$ are given in Table
~\ref{tab:SB_BP3}. 
\begin{table}[htb!]
    \centering
    \begin{tabular}{c|c|c|c|c}
    \hline\hline
    &\multicolumn{2}{c|}{Fiducial Cross section (fb)}&\multicolumn{2}{c}{Significance, $S$}\\ \cline{2-3}
       Selection   criteria  & {Signal} & {Background}&\multicolumn{2}{c}{at integrated luminosity} \\ \cline{4-5}
    (BP3) &&&300 fb$^{-1}$& 3000 fb$^{-1}$\\
       \hline\hline
       Initial (no cut) & 5.52  &506.11&&  \\
       \hline
       reject : $|y(\tau)|< 2.5$  &0.096&0&5.39&16.98  \\
       \hline
    \end{tabular}    \caption{Fiducial cross section of the signal ($pp\to H_2^+H_2^-\to \nu_\tau \nu_\tau \tau^+\tau^-$) and the SM background ($pp\to \nu\nu\tau^+\tau^-$) corresponding to {\bf BP3}  at 14 TeV LHC, with and without the selection criteria considered, and the signal significance (Eq. \ref{eq:significance}) at the HL-LHC.}
    \label{tab:SB_BP3}
\end{table}

\begin{figure}[htb!]
\centering
\includegraphics[scale=0.45]{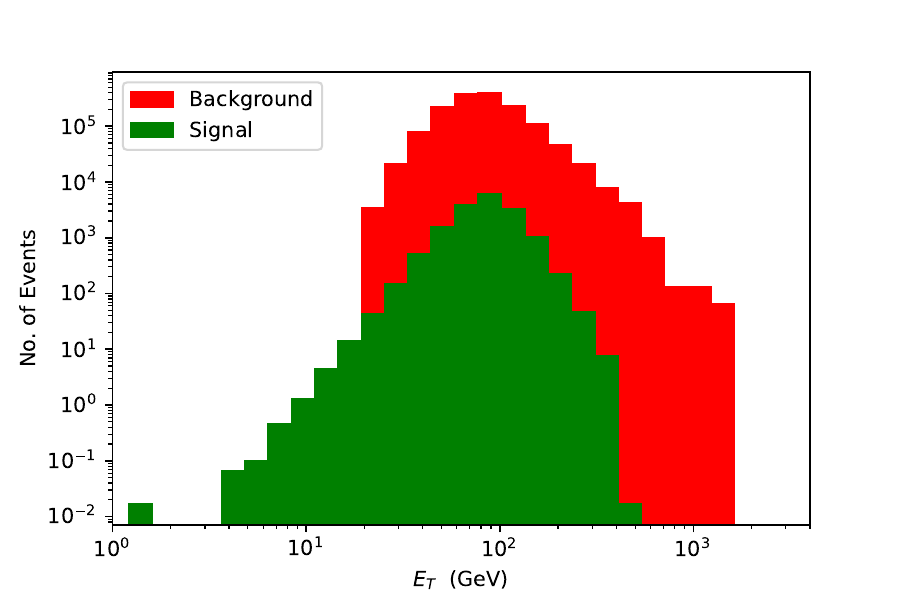}\qquad
\includegraphics[scale=0.45]{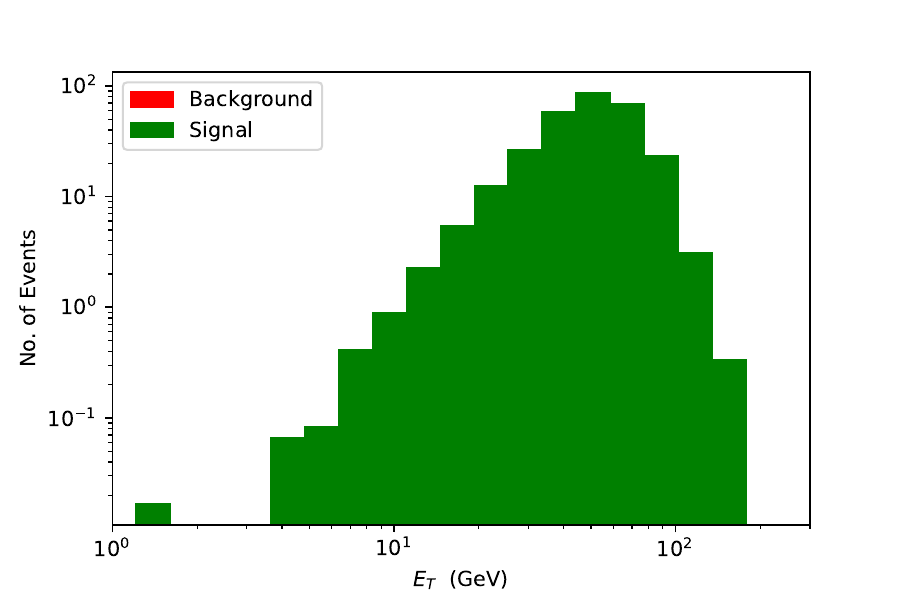}\qquad \\
\includegraphics[scale=0.45]{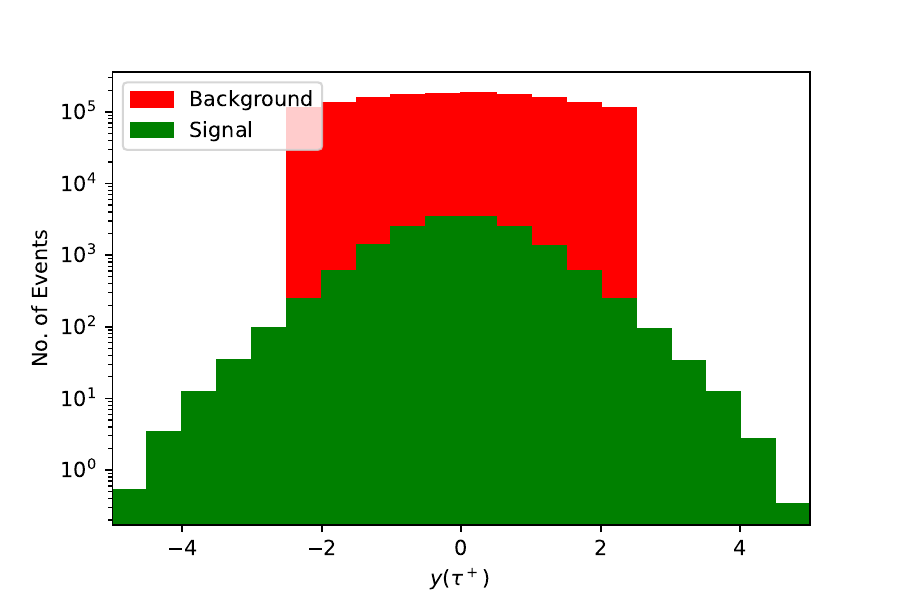}\qquad
\includegraphics[scale=0.45]{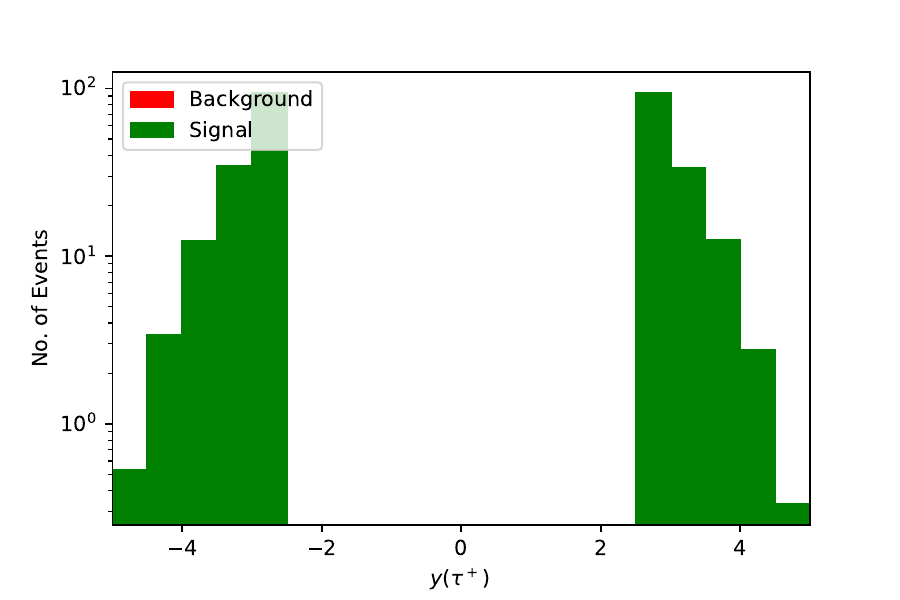}\qquad
\includegraphics[scale=0.45]{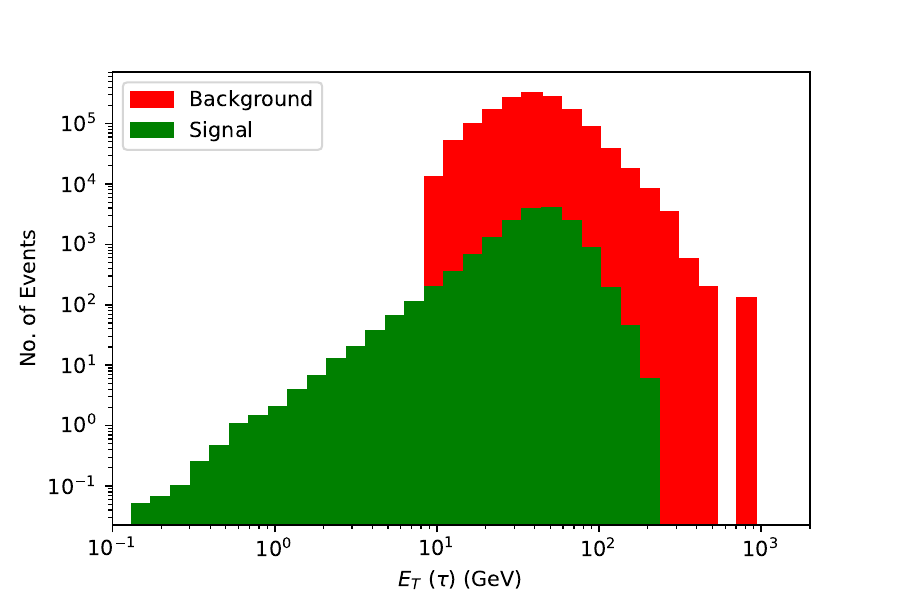}\qquad
\includegraphics[scale=0.45]{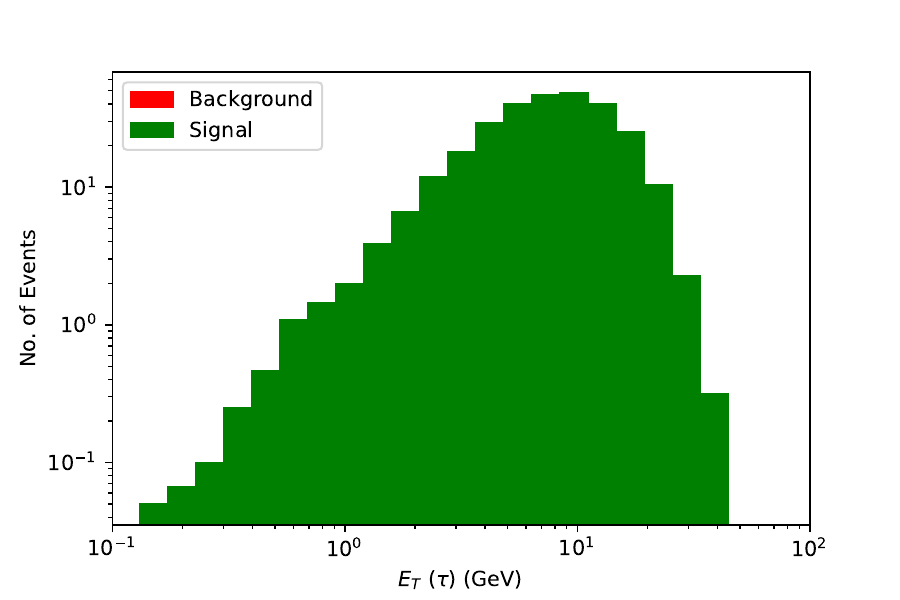}\qquad\\
\includegraphics[scale=0.45]{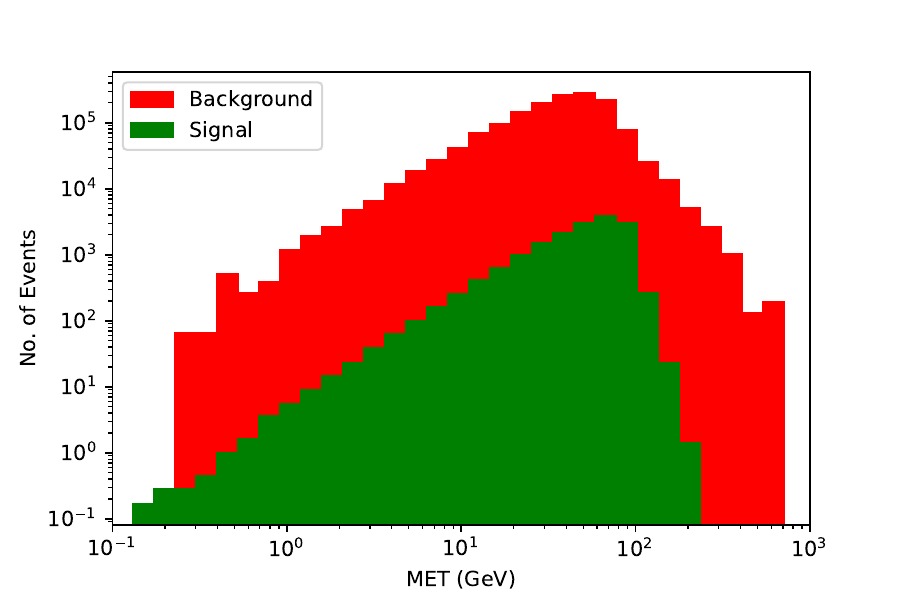}
\includegraphics[scale=0.45]{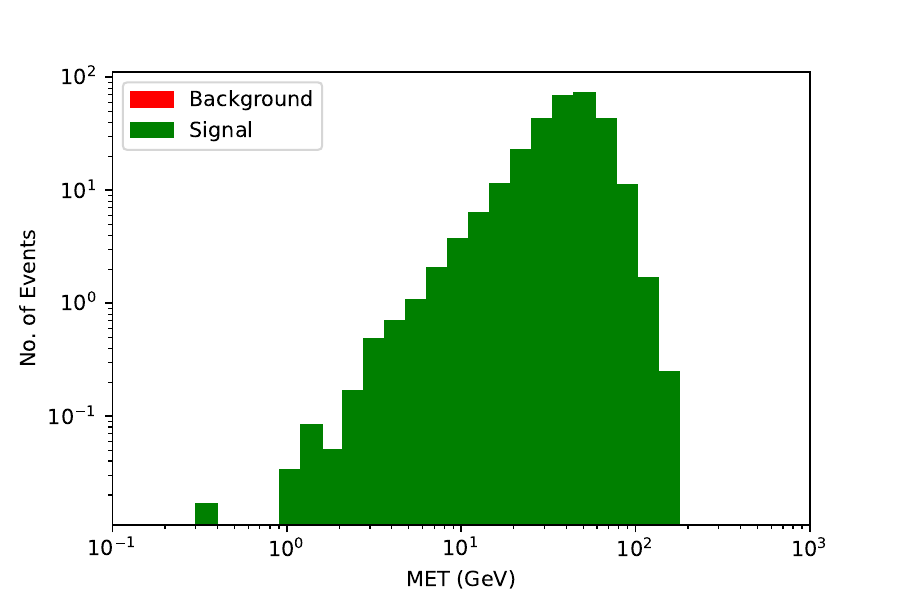}
\vskip-0.5cm
\caption{Parton level event distributions for $pp\to H_2^+H_2^-\to n_\tau n_\tau \tau \tau$ for {\bf BP3} normalized to an integrated luminosity of 3000 fb$^{-1}$ along with the SM background (in red).  
(Left column, top to bottom): the total transverse energy $E_T$, the rapidity $y$, the total transverse energy of the $\tau^+$ lepton, $E_T(\tau^+)$, and the missing transverse energy $\slashed{E}_T$, before cuts.
(Right column, top to bottom): the same distributions after the selection criteria of $|y|>2.5 $, which eliminates the background.  }
\label{fig:scotino_BP3}
\end{figure}

In Fig. \ref{fig:scotino_BP3} (left column) we show some of the kinematic distributions normalized to an integrated luminosity of 3000 fb$^{-1}$ for the benchmark {\bf BP3}. We plot, in the left column (top to bottom): the total transverse energy $E_T=\sum \limits_{\rm visible~particles}^{}|\vec{p}_T|$, the rapidity $y$, the total transverse energy of the $\tau$ lepton, $E_T(\tau)$, and the missing transverse energy 
$\slashed{E}_T=|\sum \limits_{\rm visible~particles}^{}\vec{p}_T|$.  We noticed that the signal events are mostly concentrated in  a larger  rapidity region, while the background is more or less evenly distributed in the central region, within $|y|<2.5$, so we impose the cuts as in Table \ref{tab:SB_BP3}.  We chose to highlight features of the $\tau^+$ lepton,  but similar results are obtained for the other $\tau$ lepton, $\tau^-$. Exploiting this, we employ a rejection criteria of $|y|<2.5$, to eliminate the background, while still keeping about 288 events at 3000 fb$^{-1}$ luminosity. Distributions after this event selection are given in Fig. \ref{fig:scotino_BP3} (right column).

\begin{figure}[htb!]
\centering
\includegraphics[scale=0.29]{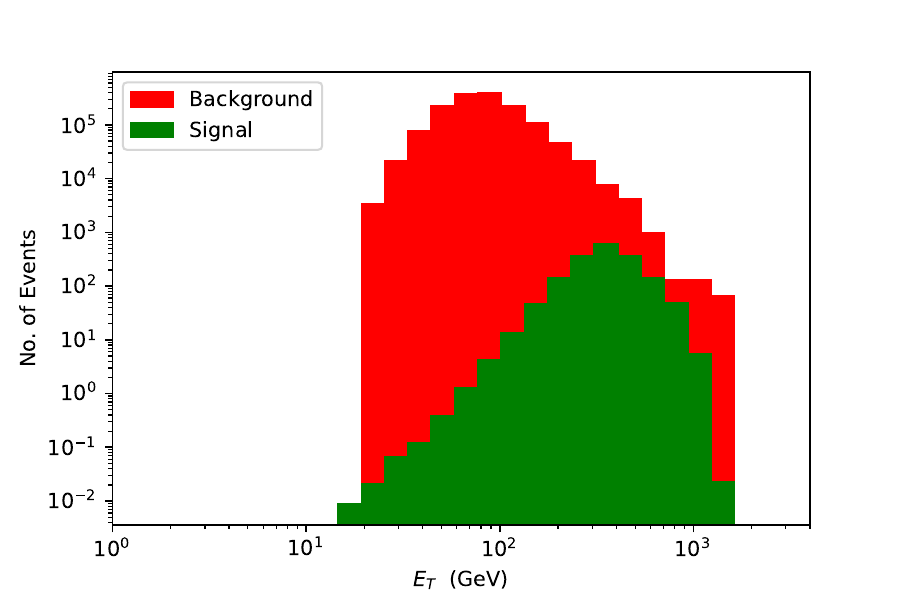}\qquad
\includegraphics[scale=0.29]{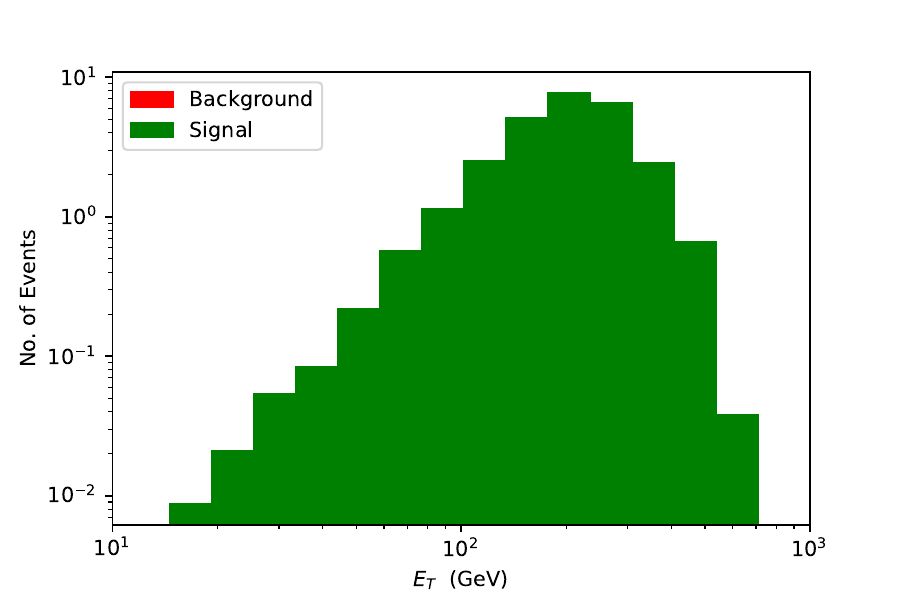}\qquad
\includegraphics[scale=0.29]{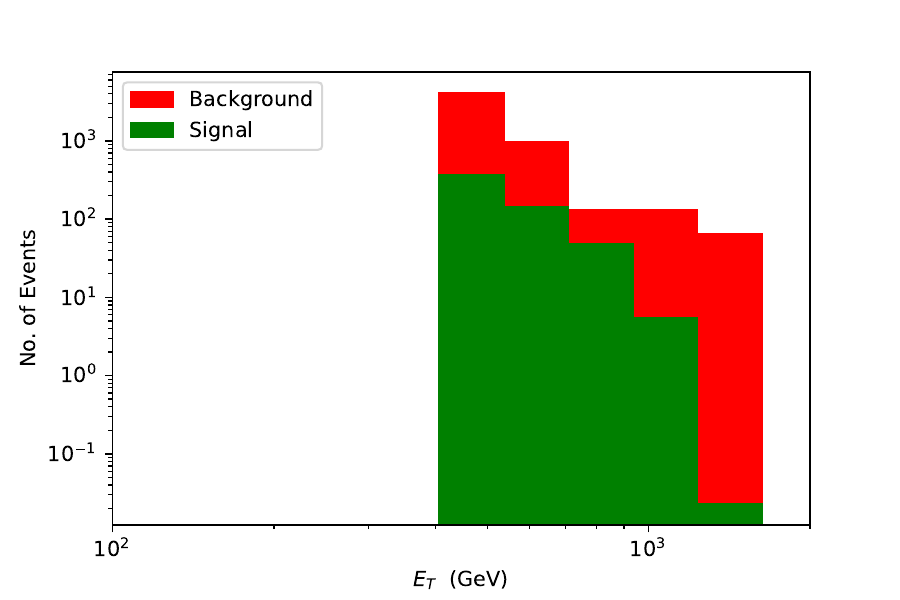}\qquad \\
\includegraphics[scale=0.29]{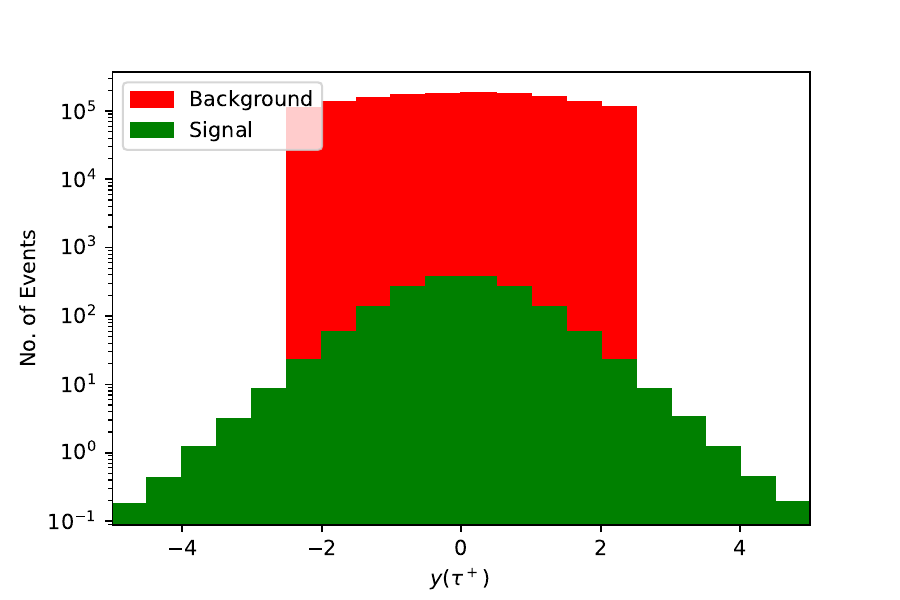}\qquad
\includegraphics[scale=0.29]{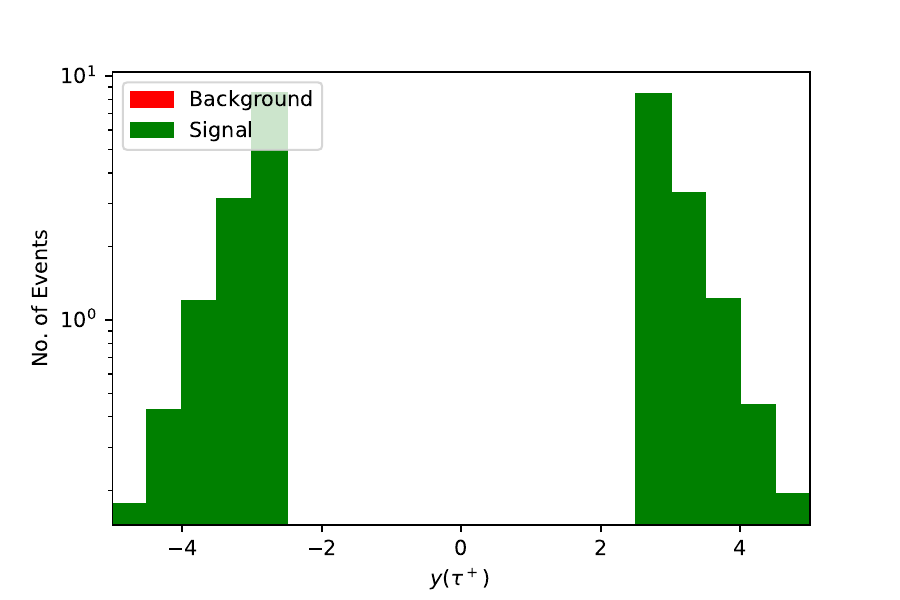}\qquad 
\includegraphics[scale=0.29]{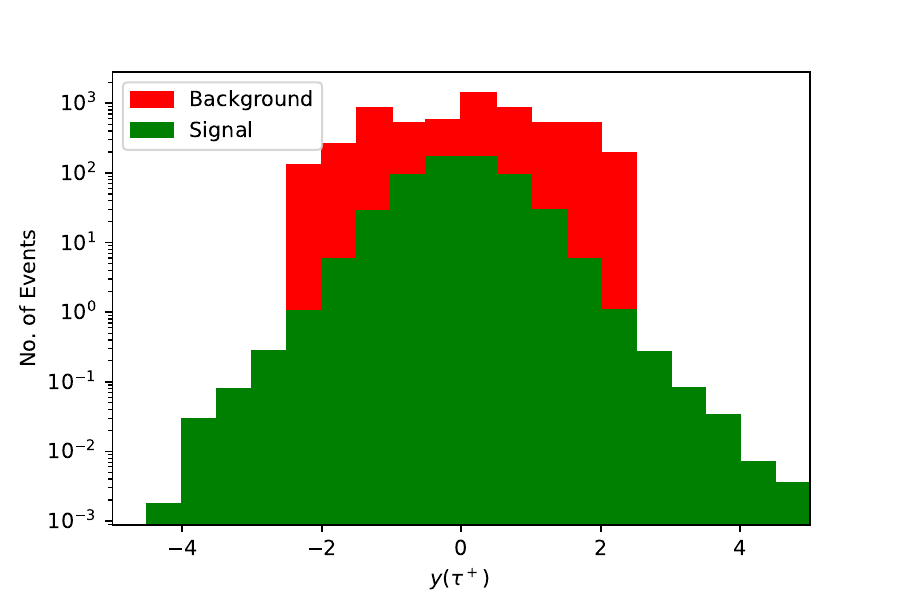}\qquad \\
\includegraphics[scale=0.29]{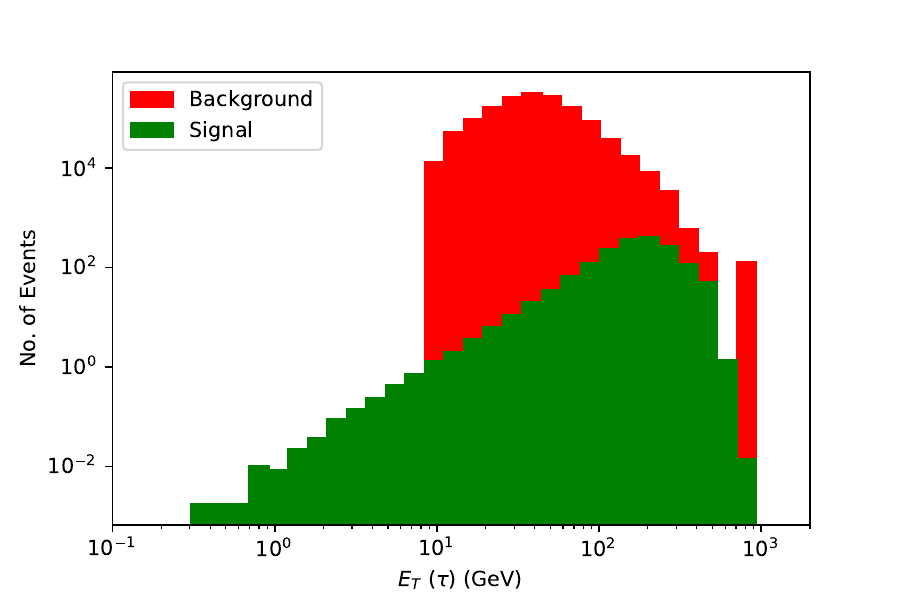}\qquad
\includegraphics[scale=0.29]{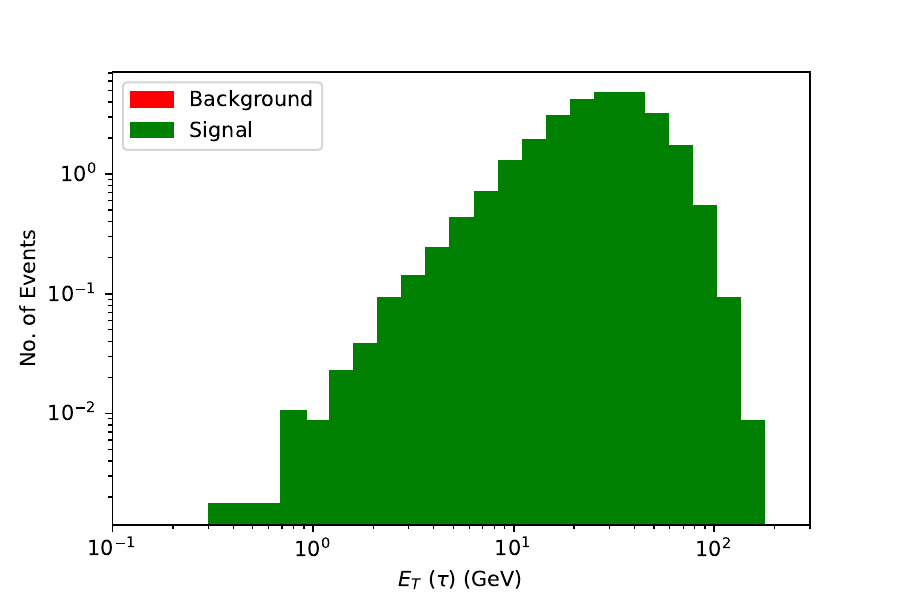}\qquad
\includegraphics[scale=0.29]{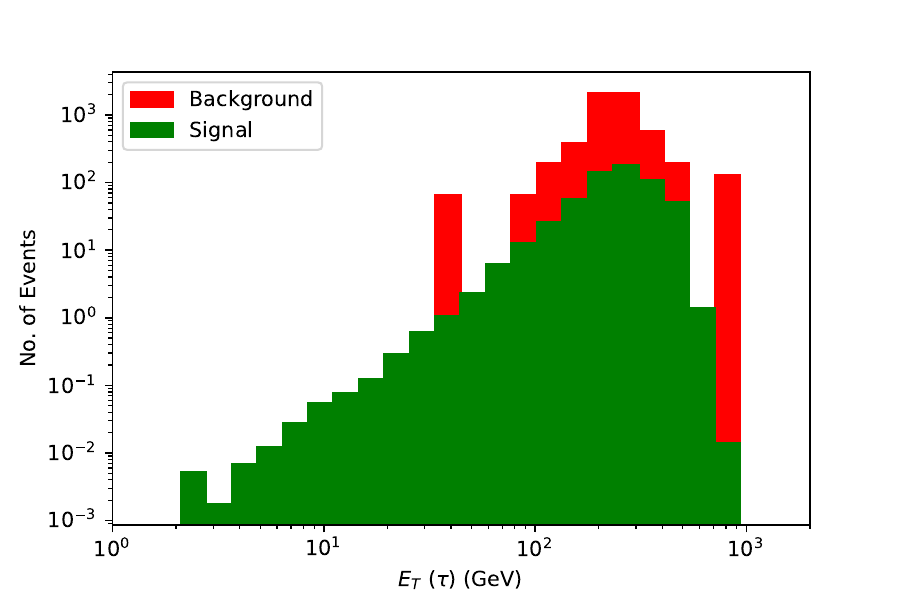}\qquad\\
\includegraphics[scale=0.29]{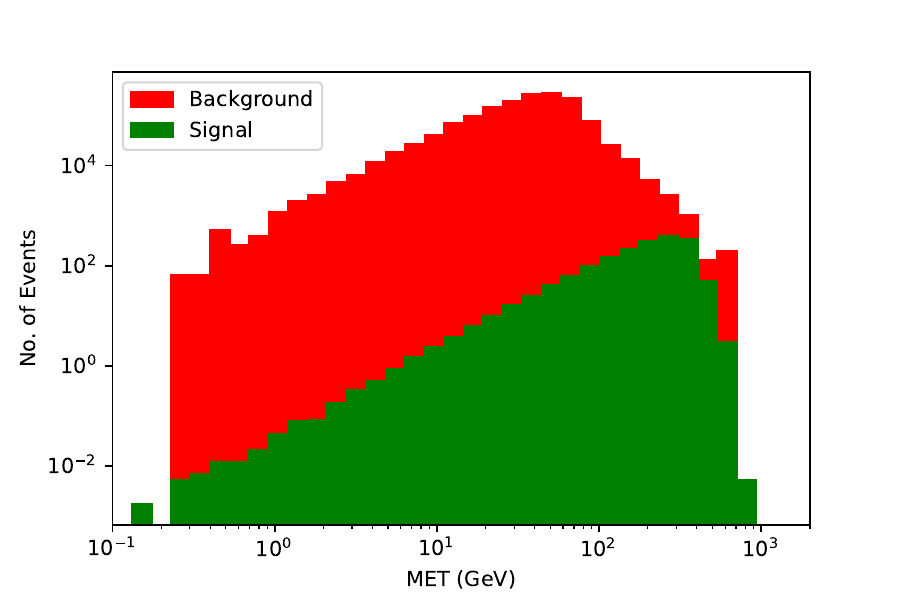} \qquad
\includegraphics[scale=0.29]{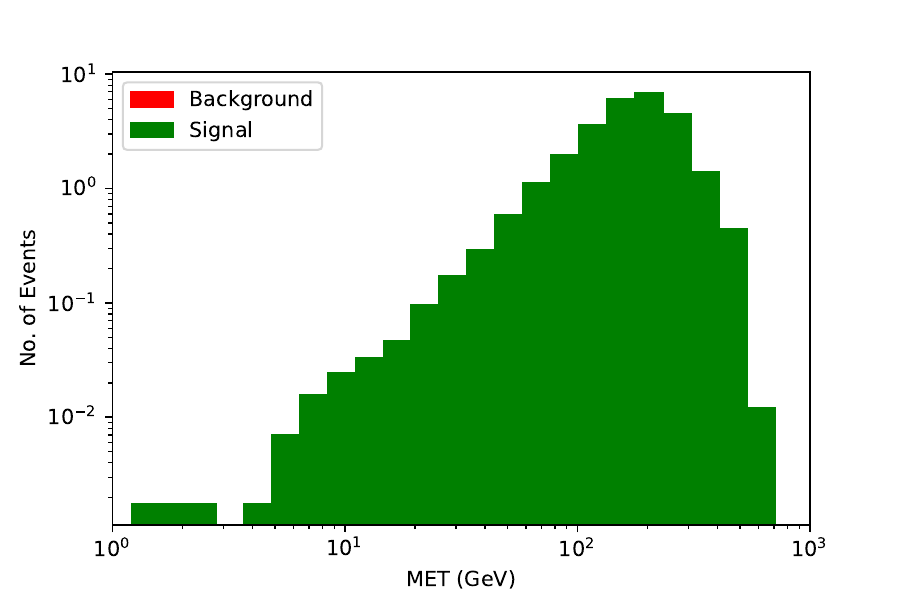} \qquad
\includegraphics[scale=0.29]{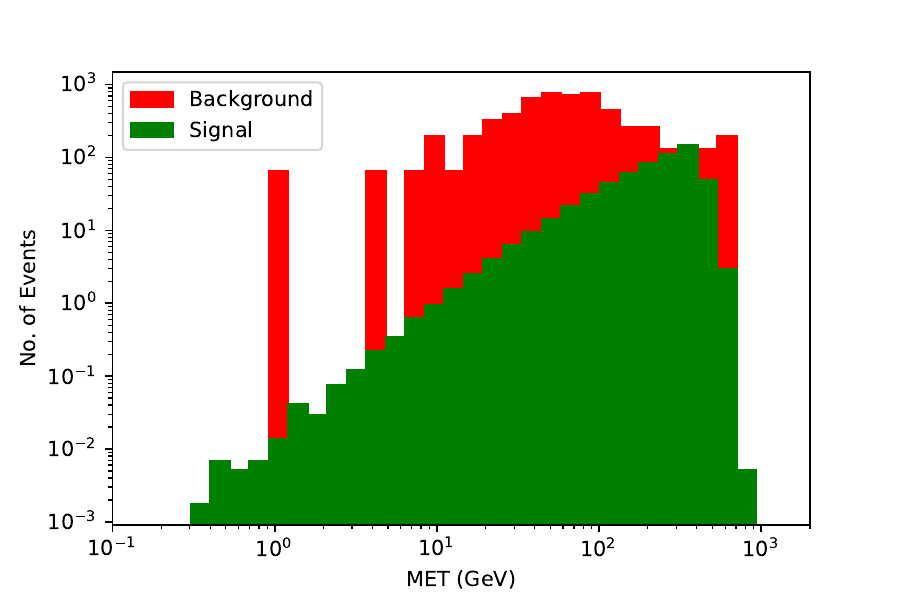} 
\vskip-0.5cm
\caption{Parton level event distributions for $pp\to H_2^+H_2^-\to n_\tau n_\tau \tau \tau$ for {\bf BP4} normalized to an integrated luminosity of 3000 fb$^{-1}$ along with the SM background (in red).  
(Left column, top to bottom): the total transverse energy $E_T$, The rapidity $y$, the total transverse energy of the $\tau^+$ lepton, $E_T(\tau^+)$, and the missing transverse energy $\slashed{E}_T$ before cuts.
(Middle column, top to bottom): the same distributions after the selection criteria of $|y|>2.5 $, which eliminates the background. (Right column, top to bottom): the same distribution, after imposing the cut $E_T>400$ GeV. }
\label{fig:scotino_BP4}
\end{figure}

The case of {\bf BP4} is similar, with the cuts given in Table \ref{tab:SB_BP4} and the event distributions given in Fig. \ref{fig:scotino_BP4}.   With the same selection criteria used in the case of {\bf BP3} (rejecting events in the central rapidity region with $|y|<2.5$) completely eliminates the background, but yields a cross section of only $0 .01$ fb and only 30 events at the highest luminosity 3000 fb$^{-1}$, with detector efficiencies estimated to be 20\%. 

Selecting an alternative cut, $E_T > 400$ GeV, the cross section is reduced to $0.2$ fb cross section for the signal after cuts, corresponding to 595 events at 3000 b$^{-1}$. While these cuts do not eliminate the background, they reduce it to 2 fb corresponding to 5834 events.  The significance is however larger than the case of rapidity cut eliminating the whole of background, 2.3 at 300 fb$^{-1}$ and 7.4 at 3000 fb$^{-1}$. We show the results for both  cuts in Table \ref{tab:SB_BP4}, and give the plots for both the rapidity and for the transverse energy cuts in Fig. \ref{fig:scotino_BP4}.

\begin{table}[htb!]
    \centering
    \begin{tabular}{c|c|c|c|c}
    \hline\hline
    &\multicolumn{2}{c|}{Fiducial Cross section (fb)}&\multicolumn{2}{c}{Significance, $S$}\\ \cline{2-3}
       Selection   criteria  & {Signal} & {Background}&\multicolumn{2}{c}{at integrated luminosity} \\ \cline{4-5}
    (BP4) &&&300 fb$^{-1}$& 3000 fb$^{-1}$\\
       \hline\hline
       Initial (no cut) & 0.58  &506.11&&  \\
       \hline
       reject : $|y(\tau)|< 2.5$  &0.009&0&1.73&5.22  \\
            \hline
       reject : $E_T < 400$ GeV  &0.198&1.945 &2.3 &7.41  \\
       \hline
    \end{tabular}    
    \caption{Fiducial cross section of the signal ($pp\to H_2^+H_2^-\to \nu_\tau \nu_\tau \tau^+\tau^-$) and the SM background ($pp\to \nu\nu\tau^+\tau^-$) corresponding to {\bf BP4}  at 14 TeV LHC, with and without two selection criteria considered, and the signal significance (Eq. \ref{eq:significance}) at the HL-LHC. }
    \label{tab:SB_BP4}
\end{table}

Note, however that observability of {\bf BP3} is more likely than of {\bf BP4} due to the larger corresponding cross section for $pp\to H_2^+H_2^-$. This yields an increase  of the number of events of {\bf BP3} compared to {\bf BP4}, and has a measurable impact on the signal significance.

This feasibility study  demonstrates that, under favourable conditions, the HL-LHC may be able probe the presence of $H_2^+$ around 1 TeV mass in this model within a few years of its run, while heavier ones need to wait for much larger integrated luminosity.  

\section{Conclusion}
\label{sec:conclusion}
We examined the dark matter candidates in ALRM and the opportunity to observe signals compatible with these choices at the LHC. Introducing an additional $U(1)_S$ symmetry allows to define  a generalized lepton number, while protecting the model against low energy flavor violation, and allows the additional Higgs bosons to be relatively light.   In this scenario an $R$-parity arises naturally, as in supersymmetry, enabling us to distinguish among $R$-parity positive particles (which include all the SM particles) and $R$-parity negative particles, the lightest of which is stable and thus a candidate for dark matter. Depending on the mass hierarchy, the model allows either a scalar dark matter (neutral $R$-parity negative Higgs boson and pseudoscalar) or a fermion (scotino) dark matter. 

We study  the parameter space compatible with the dark matter relic density, and the direct and indirect dark matter experiments for these two possibilities separately to find compatible regions. While the scalar dark matter case accommodates lighter masses, the scotino case restricts the dark matter mass to be above a TeV.  We note that while mixed scalar-fermion dark matter  could exist, to yield  the correct relic density, the scalar dark matter scenario requires the scotinos to be almost degenerate in mass with the DM candidate,  and thus the mixed case resembles closely the scalar case.  This approximate degeneracy means that  constraints from dark matter experiments require both scalar and fermionic candidates to have masses in the TeV region. We find that in all scenarios, degenerate, or small and larger mass splittings among the scotinos can be accommodated, and we  indicate the parameters sensitive to the experimental constraints and restrict the parameter space accordingly.

We follow the dark matter analysis by an exploration of the collider signature of the model. We devise four benchmarks, two for the scalar dark matter case, and two for the scotinos. The scalar dark matter could be probed by the production through two exotic quarks $pp \to d^\prime {\bar d}^\prime$, followed by the branching ratios into $b H_1^0/A_1$ (where $H_1^0/A_1$ is the DM candidate). This production-decay mode yields a  relatively large number of signals, and even though the QCD background is much larger, cuts on the rapidity and transverse energy give a promising signal over background ratio for two representative sets of model parameter values presented in {\bf{BP1}} and {\bf{BP2}}. Even considering that the background can be much larger, and with conservative detector efficiencies, this signal can be probed at LHC with 300 fb$^{-1}$, and would be indicative of exotic $d^\prime$ quarks with masses in the TeV range.

The scotino benchmarks {\bf BP3} and {\bf BP4} also prove to be  promising. Produced through $pp \rightarrow H_2^+ H_2^-\to n_\tau n_\tau \tau^+\tau^-$ (with $n_\tau$ the dark matter candidate), the process also yields events that are much smaller than the corresponding background.  However, a  cut on the rapidity in the central region renders the signal above the background, with a significance of between 1.7 to 17 at the LHC operating at 14 TeV, with luminosity of 300-3000 fb$^{-1}$. For the second benchmark {\bf BP4}, a cut on the transverse energy, while not eliminating the background, yields a better signal significance. In particular, observing {\bf BP3} is promising at 300 fb$^{-1}$, while {\bf BP4} would be observable only at 3000 fb$^{-1}$.  This signal would be indicative of the presence of charged Higgs bosons with masses the TeV region. While a more detailed collider study including realistic detector effects is beyond the scope of our work, our parton level simulations indicate the model shows some promise in  finding  signals for the lightest charged Higgs with mass in the TeV range with reasonable integrated luminosity, even accounting for detector efficiencies of 20-30\%.

\begin{acknowledgments}
The work of MF has been partly supported by NSERC through grant number SAP105354. CM and SS acknowledge the Institute Postdoctoral Fellowship of IIT Bombay where most of the work was  done. SS is partially funded for this work under the US Department of Energy contract DE-SC0011095. CM acknowledges the support from "The Scientific and Technological Research Council of Türkiye (TÜBİTAK) BİDEB 2232-A program" under project number 121C067 during his stay at METU and also thanks Ufuk Aydemir and the hospitality of METU guest house during the stay. The work of UAY is supported by an
Institute Chair Professorship. The authors are grateful to Benjamin Fuks and Olivier Mattelaer for their help to resolve some issues related to model files implementation in {\tt MadGraph}. CM and SS would also like to thank Songshaptak De for useful discussions on collider analysis.
\end{acknowledgments}
\bibliographystyle{JHEP}
\bibliography{ALRM_DM}
\end{document}